\expandafter\def\csname ver@fixltx2e.sty\endcsname{}
 \PassOptionsToPackage{pdfpagelabels=false}{hyperref} 
\documentclass[useAMS,usenatbib]{mnras}

\usepackage{aas_macros}
\usepackage{amsmath}
\usepackage{amssymb}
\usepackage{float}
\usepackage{graphicx}
\usepackage{lscape}
\usepackage{url}
\usepackage{natbib}
\usepackage{placeins}
\usepackage{color}
\usepackage{wasysym}
\usepackage{hyperref}
\usepackage{scrextend}

\newcommand{\revision}[1]{{#1}}
\newcommand{\typo}[1]{{#1}}

\begin{document}

\title{Cooling and Instabilities in Colliding Flows}

\author[R.N. Markwick et al.]%, A. Frank, J. Carroll-Nellenback, B. Liu, E. Blackman, S.V. Lebedev, P.M. Hartigan]
{
R.N. Markwick$^{1}$,  A. Frank$^{1}$, 
J. Carroll-Nellenback$^{1}$, B. Liu$^{1}$, E.G. Blackman$^{1}$, 
\newauthor 
\ S.V. Lebedev$^{2}$, P.M. Hartigan$^{3}$ \\
$^{1}$Department of Physics and Astronomy, University of Rochester, Rochester, NY\\
$^{2}$Department of Physics, Imperial College, London\\
$^{3}$Department of Physics and Astronomy, Rice University, Houston, TX
}

\date{}

\pagerange{\pageref{firstpage}--\pageref{lastpage}}
\maketitle
\label{firstpage}

\begin{abstract}
Collisional self-interactions occurring in protostellar jets give rise to strong shocks, the structure of which can be affected by radiative cooling
within the flow.
To study such colliding flows, we use the AstroBEAR AMR code to conduct hydrodynamic simulations in both one and three dimensions with a power law cooling function.
The characteristic length and time scales for cooling are temperature dependent and thus may vary as shocked gas cools.
When the cooling length decreases sufficiently rapidly the system becomes unstable to the radiative shock instability, which produces oscillations in the position of the shock front; these oscillations can be seen in both the one and three dimensional cases.
Our simulations show no evidence of the density clumping characteristic of a
thermal instability, even when the cooling function meets the expected criteria.
In the three-dimensional case, the nonlinear thin shell instability (NTSI) is found to dominate  when the cooling length is sufficiently small. When the flows are subjected to the radiative shock instability, oscillations in the size of the cooling region allow NTSI to occur at larger cooling lengths, though larger cooling lengths delay the onset of NTSI by increasing the oscillation period.
\end{abstract}

\begin{keywords}
Herbig–Haro objects -- hydrodynamics  --  ISM: jets and outflows -- instabilities -- methods: numerical -- shock waves
\end{keywords}

\section{Introduction}
Hypersonic flows of magnetized plasmas naturally produce strong shocks when obstructions or self-interactions (flow collisions) occur. Such complex flows occur in a wide variety of astrophysical settings.  These include supernova explosions, shock cloud interactions, space weather and cloud collisions during galaxy interactions.  Colliding flows also occur in High Energy Density Plasma (HEDP) settings as well such as Z-pinches and laser driven implosion experiments. 

 A particularly noteworthy form of colliding flows are the highly collimated outflows known as  "protostellar jets" or "Herbig Haro jets" \citep{Frankea2014}.
These occur when a young stellar object (YSO) ejects matter via rotational magnetohydrodynamic processes to form a dense and narrow beam of collimated matter \citep{rayPPV2007, Frankea2014}.  
The colliding flow aspect of these jets comes from observations which often show chains of knots \citep{Hartigan11}. This morphology likely arises from collisions between faster moving jet material with slower material ejected earlier. Hydrodynamic and MHD evolution in these interaction zones produce is then expected to lead to high degrees of heterogenity or "clumpiness" \citep{Hansen17}. 

Radiative cooling will be significant for jet systems in star forming regions.  Over the years many simulations have been performed to study various aspects of "radiative jets" such as basic flow properties \citet{dalPino96} the role of magnetic fields (\citet{Gardiner00,Osullivan00} and the interactions between jet flows and protostellar disc winds. 

All supersonic jets, whether adiabatic or cooling due to optically thin radiative losses, are structurally composed of a supersonic beam, a cocoon of shocked jet gas at the end of the beam, a region of shocked ambient gas pushed by the cocoon, and a bow shock \citep{blondin90}. The bow shock moves at velocity \citep{DalPino93}
\begin{equation} 
v_{bs} = v_j \left[1+(\eta\alpha)^{-\frac12}\right]^{-1},
\end{equation}
where $v_{bs}$ and $v_j$ are the respective velocities of the bow shock and the jet, $\eta$ is the ratio density of the jet material to that of the ambient medium ($\rho_j/\rho_a$), and $\alpha$ is the square of the ratio of the radius of the beam to the radius of the head of the jet. Note that for radiative jets a number of dimensionless parameters are used to characterize the system, such as $\eta$ defined above, the mach number $M$, and the cooling parameter $\chi = t_\text{cool}/t_{hydro}$, which is given as the ratio between cooling time-scale (defined in section \ref{sec:theory_scales}) and hydrodynamic time-scale.

When radiative energy losses are significant ($\chi\lesssim 1$),  jets tend to exhibit instabilities occurring in the regions behind shocks. In particular, three dimensional models show considerable structure formation occurring in unstable regions associated with shocks, leading to strong "clumping" in the simulated jets much like what is seen in the observation.  

The mode involved in these instabilities is, however, not always clear. There are a number of unstable modes available in radiatively cooling regions in protostellar jets.  First there is the {\it radiative shock instability} which leads to oscillations in the shock front \citep[e.g.][]{Langer81,stricklandBlondin95,Walder98}. The {\it Nonlinear Thin Shell Instability} (NTSI) \citep{Vishniac94}, which is a bending mode of the "cold slab" of post-shock cooled material can also occur in protostellar jets. Finally, there is the {\it Thermal} or {\it Field} instability \citep{field65,Balbus86} in which cold clumps condense out of a hotter background. 

The question of which instability dominates in radiatively cooling jets also comes up in laboratory experiments, which have been used to study the properties of plasma jets. While the physical scale of laboratory experiments differs from that of astrophysical jets by several orders of magnitude,
correspondence of results can be obtained through the use of dimensionless parameters as established in \citet{ryutov2000}.  Experiments performed at Imperial College \citep{suzukiVidal09, Ciardi} and Cornell University \citep{gourdain10} have created single radiatively cooling jets using a pulse-powered generator driving a radial foil z-pinch. Collimation occurs via the formation of a magnetic tower of toroidal fields as predicted by \citet{LyndenBell96}. In \citet{suzukiVidal12} the interaction of a jet with the ambient medium was studied. 

Of particular interest for this study are experiments by \citet{suzukiVidal15} who examined the structure of bow shocks formed by the collision of two jet flows. Among the most important results of these experiments was the formation of small-scale structures in the interaction region where the two jets collided. The time-scale over which the formation of these structures occurred was consistent with the estimated time-scales for radiative cooling.  In that paper the authors concluded that the most likely unstable mode was due to the thermal or "Field" instability \citep{field65} however this conjecture could not be tested.  

In this paper, the first in a series, we begin a study of the colliding radiative flows like those of \citet{suzukiVidal15}. \revision{ While astrophysical jets are likely magnetic, it is 
 prudent to build a realistic jet model in stages by understanding first the radiative hydrodynamic case, which is itself rich
in the underlying physics; we will ultimately have a better understanding of the particular signatures of  the magnetic case if we first understand the case without magnetic fields for future comparison. Our goal in the present paper is therefore to focus solely on hydrodynamic simulations using an analytic form of radiative cooling (i.e. a power-law cooling function)}, which allows us to mimic different regions of a full cooling curve. We begin with one-dimensional simulations to reproduce and extend previous studies by \citet{stricklandBlondin95} and to test our model. We then move on to three-dimensional simulations aimed to further investigate the results of \citet{suzukiVidal15} with the goal of gaining a better understanding of which instability, driven by radiative cooling behind the shock, is at work in both the laboratory experiments and astrophysical radiative jets.

This paper is organized as follows: In section \ref{sec:theory}, we begin with necessary background on the structure of colliding flows (\ref{sec:theory_morphology}) before discussing theoretical aspects of cooling (\ref{sec:theory_scales}) and instabilities (\ref{sec:theory_insta}) which may be applicable. 
In Section \ref{sec:meth} we discuss the model system and simulation parameters. 
In section \ref{sec:result}, we present the results of the simulations, with section \ref{sec:results_1D} containing the results of the 1-dimensional simulations while section \ref{sec:results_3D} contains the results of the 3-dimensional simulations. Section \ref{sec:discuss} will include a discussion of instabilities relevant to these results.

\section{Theoretical Background} \label{sec:theory}

\subsection{Morphology of Colliding Jets} \label{sec:theory_morphology}

\begin{figure}
	\includegraphics[width=\columnwidth]{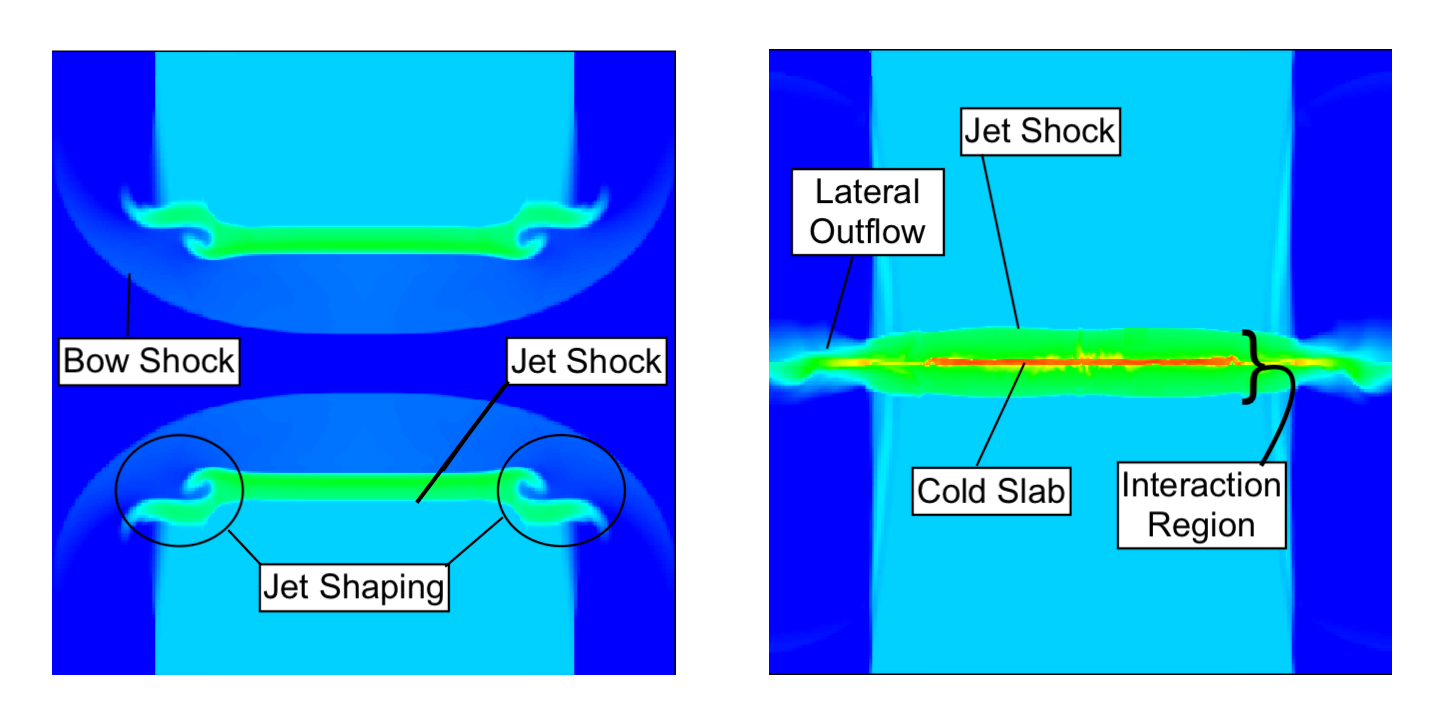}
  \caption{
Identification of key components of the colliding jet systems before and after collision.
    }
    \label{fig:3DSchematic}
\end{figure}

We begin with a brief description of the general morphology of the colliding jets in our simulations (see figure \ref{fig:3DSchematic}). Prior to collisions, the jets propagate towards each other creating the well characterized structure composed of a forward (relative to the flow direction) facing bow shock and a rearward facing jet shock. As jet material passes through the jet shock it flows sideways and away from the jet head to form a cocoon surrounding the body of the jet \citep{blondin90}. Note that Figure \ref{fig:3DSchematic} shows that the flow of post jet shock material produces feedback on the jet head in our simulations, beveling the edge of the jet. This pre-collision shaping of the jet will leave an imprint which will appear later as instabilities grow. 

After the collision of the two jets, the jet shocks become boundaries forming an {\it interaction region}.
The structure of radiative shocks always shows high temperatures directly behind the shock followed by a {\it cooling region}, in which the shocked gas goes from its post-shock temperature $T_s$ back to a lower temperature. 
Behind the cooling region is a {\it cold slab} where gas temperatures reach their final post-cooling value and densities are highest. 
Since the jets in our simulation have identical parameters, the interaction region does not change position unless disrupted by instabilities. Some shocked material gets ejected laterally by the high pressures throughout the interaction region \citep{Raga93}.

\subsection{Cooling Scales} \label{sec:theory_scales}

The impact of radiative cooling on the structure of the jet can be described \typo{in terms of the cooling length} and cooling time, which are given by \citep{DalPino93, blondin90}
\begin{equation} d_\text{cool} \approx \frac{v_s t_\text{cool}}{4} = \frac{v_s}{4}\frac{3kT_s}{n\Lambda(T)}\label{eq:dcool},\end{equation}
where $n\Lambda(T)$ is a cooling function, which gives the energy loss per unit volume, and $\frac{v_s}{4}$ gives the approximate postshock velocity for a strong \revision{$\gamma=\frac53$} shock.
Only when the cooling length is much larger than any other length scale relevant to structure formation can the system be approximated as adiabatic. Conversely a very short cooling length allows for an isothermal approximation.

In real systems "cooling functions" are based on microphysical processes usually associated with the collisional excitation and radiative de-excitation of atomic or molecular transitions. Such cooling functions can have many local minima and maxima as a function of temperature.  In this paper we wish to focus only on the slope of the cooling curves and its effect on the dynamics.  Thus the cooling function we use is a simple power law 
\begin{equation}\Lambda(T) = \alpha \left(\frac{T}{T_0}\right)^\beta, \end{equation}
where we treat $\alpha$, $\beta$ and $T_0$ as free parameters.  $\alpha$ is used to set the absolute value of the cooling strength and, hence, the immediate post-shock cooling time
$\hat{t}_\text{cool}$ is given as
\begin{equation}
  \hat{t}_\text{cool} = \left(t_\text{cool}\right)_{T=T_s} = \frac{3kT_s}{n\alpha}\left(\frac{T_0}{T_s}\right)^\beta\label{eq:tcool}.
\end{equation}
In practice we set $T_0$ to be the temperature just behind the shocks in an effort to keep $\hat{t}_\text{cool}$ consistent across a range of $\beta$.  
This value for the cooling time, along with the corresponding cooling length, 
are accurate in the region immediately behind the shock front. 

As the gas cools behind the shock, the decreasing temperature and corresponding change in density result in the local value of the cooling time changing across the cooling region. If density scales as $n = n_0 \big(\frac{T}{T_0}\big)^{\xi}$ then the cooling time scales as 
\begin{equation}
t_\text{cool} = \frac{3k T}{n\alpha}\left(\frac{T_0}{T}\right)^{(\beta)} = \frac{3k T_0}{n_0\alpha} \left(\frac{T}{T_0}\right)^{(1-\beta-\xi)} \label{eq:tcoolloc},
\end{equation} 
so the cooling time decreases as the gas cools if
\begin{equation} 
\frac{\partial(\ln\Lambda)}{\partial(\ln T)} = \beta < 1-\xi.
\end{equation}
 In a 1D flow, such as that behind a shock, conservation of mass implies that velocity changes inversely with density.  Thus the cooling length decreases as the gas cools if
\begin{equation} 
\frac{\partial(\ln\Lambda)}{\partial(\ln T)} = \beta < 1-2\xi. 
\end{equation}
 For isochoric cooling, $\xi=0$ so both the cooling length and time scales would decrease as the gas cools if $\beta <1$. For isobaric cooling, $\xi=-1$ so the cooling length decreases as the gas cools whenever $\beta<3$ while the cooling time only decreases as the gas cools if $\beta<2$. 

If $\zeta = \xi+\beta - 1$ vanishes, then the cooling time is uniform with the value given by equation \ref{eq:tcool}. The time required for cooling from  temperature $T_s$ to temperature $T_f$ can therefore be given as
\begin{equation}
t(T_f)-t(T_s) = C_t(T_f) \hat t_\text{cool} =  \ln\left(\frac{T_s}{T_f}\right) \hat t_\text{cool}.
\end{equation}
 For all other values of $\zeta$, the cooling time varies with temperature. A more accurate estimation of the time required for cooling from temperature $T_s$ to temperature $T_f$ can be estimated by multiplying $\hat t_\text{cool}$ \typo{by a correction} factor
\begin{equation}\label{eq:correction} 
C_t(T_f) = \int_{T_f}^{T_s} \left(\frac{T_0}{T}\right)^{\zeta+1} \frac{\operatorname{d} T}{T_0} = \frac{ \left(\frac{T_0}{T_f}\right)^{\zeta}-\left(\frac{T_0}{T_s}\right)^{\zeta}}{\zeta}.
\end{equation}
 In the limit where $\zeta$ approaches zero this correction factor approaches $\ln\left(\frac{T_s}{T_f}\right)$ as is expected for a uniform cooling time.
The cooling distance 
can be corrected in a similar manner except that $\zeta = 2\xi+\beta-1$. 

An additional correction to equation \ref{eq:tcool} is needed \typo{to account for the tendency} for temperature to increase as a result of compression. Written in terms of temperature, energy conservation (equation \ref{eq:Eu3}) for an ideal gas is given by
\begin{equation} \label{eq:temp}
    \frac{1}{\gamma-1}\left[\frac{\partial }{\partial t} + {\bf v}\cdot\nabla\right] T = \frac{T}{\rho}\left[\frac{\partial }{\partial t} + {\bf v}\cdot\nabla\right]\rho - n\Lambda(T).
\end{equation}
Equation \ref{eq:tcool} is correct for $\gamma=\frac53$ in the case where $\frac{\operatorname{D} \rho}{\operatorname{D} t}$ vanishes. If instead $\rho$ scales as $T^\xi$, then equation \ref{eq:temp} becomes
\begin{equation}
    \frac{1}{\gamma-1}\left[\frac{\partial }{\partial t} + {\bf v}\cdot\nabla\right] T =\xi\left[\frac{\partial }{\partial t} + {\bf v}\cdot\nabla\right] T - n\Lambda(T).
\end{equation}
 This is equivalent to the case where $\frac{\operatorname{D} \rho}{\operatorname{D} t}$ vanishes if $\Lambda(T)$ (in the case of a power law, $\alpha$) is divided by the correction factor $ 1-(\gamma-1)\xi $. Unlike the correction given by equation \ref{eq:correction}, this correction is independent of temperature and (provided that a power law relation between density and temperature holds) is uniform throughout the cooling region.

\subsection{Instabilities} \label{sec:theory_insta}

As discussed in the introduction radiative shocks are susceptible to a number of instabilities.  In this section we briefly describe the unstable modes which have been considered to be the most important based on previous studies.  These are the modes we will focus on in our study.

 The so-called radiative shock instability, first described by \citet{Langer81}, 
arises when discrepancies between the cooling length and the size of the cooling region lead to oscillations in the position of the shock front. 

When the size of the cooling region is less than the cooling length, the cooling region becomes over-pressured as a result of the post-shock gas not having sufficient time to cool before reaching the cold slab; this causes the shock to expand forward.
Meanwhile if the cooling region becomes larger than the cooling length, the shock becomes under-pressured and retreats backwards. 

\typo{In the former scenario} as the shock velocity
increases in an attempt to recover pressure equilibrium, the post-shock temperature will similarly increase; for systems with sufficiently low $\beta$ 
this will result in a decreased cooling length. Thus the efforts of the system to restore pressure equilibrium result instead in a further imbalance. While \typo{the size of cooling region} will eventually exceed the cooling length, by that point the system will have been driven beyond its steady-state configuration, resulting in oscillatory behaviour. 
 As $\beta$ is increased the oscillations become damped as the   correlation between shock velocity and cooling time changes from positive to negative \citep{stricklandBlondin95}.
 The period of oscillation is proportional to the thickness of the cooling region divided by the velocity of the preshock gas \citep{CI}, therefore decreasing as cooling strength ($\alpha$) in creases.

Theoretical work by \citet{Langer81} suggested that shocks with $\beta \gtrsim 1.1$ are stable against oscillations. \revision{
The specific critical values however may be lower and are sensitive to boundary conditions \citep{stricklandBlondin95,Mignone05}. For example, lower overtones more easily damped in the presence of a reflecting wall, since a wave traversing the cold slab can be reflected out of phase with the original shock wave; under such conditions \citet{stricklandBlondin95} found that oscillations begin to become damped at lower values of $\beta \gtrsim 0.7$ for overtones and $\beta \gtrsim0.4$ for the fundamental.
In addition to the effects of boundary conditions, critical values are also found to decrease at lower specific heat ratio $\gamma$ \citep{Ramachandran05} and at lower Mach numbers \citep{Ramachandran06, Pittard}.}
 Multi-dimensional simulations  \citep{sutherland03b} demonstrate that pulsations in the direction of the shock still occur though the flow becomes turbulent which can lead to the development of structure within the cooling zone and cold slab (see figure 12 of \citet{sutherland03b}).

A second instability relevant to our simulations is the nonlinear thin shell instability (NTSI) \citep{Vishniac94, BlondinMarks96}. This instability arises from the interaction between ram pressure of the incoming flow and the thermal pressure inside the shell (see figure 1 of \cite{McLeod13}). In unperturbed conditions, the direction of the flow is normal to the surface of the shell, so the gradients of these pressures cancel as a result of this alignment.  If however the shell is perturbed by a finite amount, the surface of the shell can become misaligned from the flow.
Such a misalignment results in an imbalance of forces, pushing additional shell material away from the regions of greatest misalignment and toward the regions of greatest perturbation zones; this promotes growth of the perturbation. The growth rate of perturbations of amplitude \revision{$\psi$} and wavelength $\lambda$ is given by
\revision{\begin{equation} \label{eq:NTSI}
\dot \psi \sim \left(\frac{\psi}{\lambda}\right)^{\frac32} c_s.
\end{equation} }
The thickness of the shell limits the NTSI in two ways. First, perturbations are stable to linear order and require a seed of thickness greater than that of the shell. Second, corrugation of the slab is limited to wavelengths larger than the thickness of the shell; longer wavelengths have slower growth rates per equation \ref{eq:NTSI} and thus the maximum growth rate for NTSI is slower for thicker shells
\citep{McLeod13}.
\revision{The effect of shell thickness also plays a role in saturation of the NTSI: \citet{BlondinMarks96} found that as perturbations grow, the shell increases in width, eventually exceeding one half-wavelength and inhibiting further growth. 
 }

 \revision{Instabilities such as the NTSI do not occur in isolation, rather other effects can promote, inhibit, or result from the instability. 
 Some instabilities such as the transverse acceleration instability may exist alongside the NTSI \citep{Dgani96}, though in the isothermal limit the NTSI is found to dominate the long-term evolution of the shock front even  when other instabilities have higher estimated growth rates \citep{blondin98,Lamberts11}.
Other} instabilities \typo{which can occur} in our set-up may act as seeds for the NTSI. When two collimated flows collide, turbulence can be seen to arise in the cold slab and cooling region as a result of either Rayleigh-Taylor and Richtmyer-Meshkov \citep{Walder98}. Velocity shear, \revision{which arises from the obliquity of the slab and promotes growth of the NTSI \citep{BlondinMarks96}}, can also trigger turbulence via the Kelvin-Helmholtz instability \citep{Stevens92, Lamberts11}.  Thus turbulent behaviour in the cold slab may result in bending modes that trigger the NTSI \citep{Folini06}. The spatial scale of these structures grows with time as the cold slab accumulates matter. 
\revision{ Finally, as obliquity of the shock increases, the shock temperature falls, resulting in reduction in thermal X-Ray emissions \citep{Steinberg18}.}
 
The final instability we focus on is the {\it Thermal} or {\it Field} instability described by \citep{field65}. This mode is applicable in regions of the fluid lying behind the shock where cooling occurs. The tendencies of a fluid to evolve towards pressure equilibrium results in the compression of cooler regions. Thus, in the presence of cooling, pockets of higher density and lower temperature than their surroundings can form.  For instability to occur, the cooling time must decrease with decreasing entropy \citep{Balbus86}. Since entropy decreases as the gas cools, the instability criterion can be expressed as 
\begin{equation} 
\frac{\partial(\ln\Lambda)}{\partial (\ln T)} < \begin{cases} 
+1.0 & \text{isochoric flow} \\ +2.0 & \text{isobaric flow}\end{cases}.
\end{equation}
 Isobaric modes are unstable for higher values of $\beta$ since the increase in density of an isobaric perturbation provides an increase in the cooling rate, counteracting the reduction in cooling arising from decrease in temperature.
\revision{More recently, \citet{Falle2020} reanalysed the stability criteria and obtained the following results: first, isobaric instability occurs if  in equilibrium pressure decreases with increasing density; second, isentropic instability occurs if the equilibrium sound speed exceeds the frozen (adiabatic) sound speed; third, that the inclusion of magnetic fields does not alter the stability criteria. }

\section{Methods and Model} \label{sec:meth}

The simulations in this study were conducted using AstroBEAR\footnote{https://astrobear.pas.rochester.edu/} \citep{cunningham09,carroll13}, which is a massively parallelized adaptive mesh refinement (AMR) code that includes a variety of multiphysics solvers, such as magnetic resistivity, radiative transport, self-gravity, heat conduction, and ionization dynamics.  Our study uses only the hydro solvers with an energy source term associated with the radiative cooling (see section \ref{sec:theory_scales}). Thus our governing equations are:
\begin{equation}
    \frac{\partial \rho}{\partial t} + \boldsymbol{\nabla} \cdot \rho \boldsymbol{v} = 0 
    \label{eq:Eu1}
\end{equation}
\begin{equation}
    \frac{\partial \rho \boldsymbol{v}}{\partial t} + \boldsymbol{\nabla} \cdot \left ( \rho \boldsymbol{v} \otimes \boldsymbol{v} \right )= - \nabla p
    \label{eq:Eu2}
\end{equation}
\begin{equation} 
    \frac{\partial E}{\partial t} + \boldsymbol{\nabla} \cdot ((E + p) \boldsymbol{v}) =-n^2\Lambda(T)
    \label{eq:Eu3}
\end{equation}
where $\rho$ is the mass density, $n$ is the number density of nuclei, $\boldsymbol{v}$ is the fluid velocity, $p$ is the thermal pressure, $\phi$ is the gravitational potential, and $E = \frac{1}{\gamma - 1} p + \frac{1}{2}\rho v^2$ is the combined internal and kinetic energies. \revision{In all runs an average particle mass of 1 amu was used.}

\revision{Cooling was applied only at temperatures above the floor temperature of the simulation in order to safeguard against runaway cooling. Since realistic cooling curves tend to vanish at low temperatures, this behaviour is justifiable on physical grounds, though it may be worth noting that \citet{Pittard} found that cooling below the pre-shock temperature enhances the radiative shock instability.}

For the one dimensional runs we do not run two jets into each other. Rather we use boundary conditions with extrapolation \revision{(specifically, Neumann boundary conditions with a derivative of zero)} on the left and a reflecting wall on the right. The flow (the jet) is sent into the grid from the left boundary and a shock is established when this flow  collides with the reflecting wall on the right boundary. The physical scales of flow where chosen to be applicable to the laboratory experiments of \citep{suzukiVidal15}. The dimensionless numbers for our simulations ($M, \eta, \chi$) place them in the same family of flows as  protostellar jets. The flow is injected with a uniform density of $10^{17}$ particles per cm${}^3$, a velocity of +47 km s${}^{-1}$, and a temperature of 720K. $\alpha$ and $T_0$ were fixed at $8.27\times 10^{-23} $ erg cm${}^{3}$ s${}^{-1}$ and $5.060\times 10^4$ K respectively. 
$\beta$ was varied across a range from $-1.0$ to 3.0 in half-integer increments. For temperatures close to $T_0$, these parameters give a cooling time of $\hat{t}_\text{cool} = 6.33\times 10^{-7}$ s and a cooling length of $\hat{d}_\text{cool} = 2.97$ cm.

 \begin{table}
 \revision{
     \centering
      \begin{tabular}{|c||c|c|c|c|c|}\hline
    figure & radius & $\alpha$ & $\beta$  & resolution  & $d_\text{cool}$\\ \hline\hline
   \ref{fig:3DT1},\ref{fig:3DS2} & 2.0 cu & $\alpha_0$  & -1.0 & 0.0625 cu & 1.50 cu\\\hline
  \ref{fig:3DS2} & 2.0 cu & $\alpha_0$  & 0.0& 0.0625 cu& 2.00 cu\\\hline
    \ref{fig:3DS2} & 2.0 cu & $\alpha_0$  & +1.0& 0.0625 cu&2.97 cu \\\hline
  \ref{fig:3DS2} & 2.0 cu & $\alpha_0$  & +2.0& 0.0625 cu& 5.41 cu  \\\hline
  \ref{fig:3DT3},\ref{fig:3DS2} & 2.0 cu & $\alpha_0$  & +3.0& 0.0625 cu& 10.36 cu  \\\hline
  \ref{fig:3DL1}, \ref{fig:3DOsc} & 16.0 cu & $\alpha_0$ & -1.0 & 0.125 cu & 1.50 cu  \\\hline
      \ref{fig:3DL2} & 16.0 cu & $\alpha_0$&+1.0 &  0.125 cu  & 2.97 cu \\\hline
 \ref{fig:3DL3} & 16.0 cu & $\alpha_0$ & +3.0 &  0.125 cu & 10.36 cu \\\hline
  \ref{fig:3DLC1} &16.0 cu & $10\alpha_0$ & -1.0& 0.125 cu  & 0.15 cu \\\hline
    \ref{fig:3DLC2} & 16.0 cu & $10\alpha_0$ & +1.0 & 0.125 cu & 0.30 cu \\\hline
    \ref{fig:3DLC3} & 16.0 cu & $10\alpha_0$ & +3.0 & 0.125 cu & 1.04 cu  \\\hline
      \ref{fig:3DLH1} & 16.0 cu & $0.1\alpha_0$ &-1.0& 0.125 cu  & 15.0 cu \\\hline
     \ref{fig:3DLU3} & 16.0 cu & $100\alpha_0$ & +3.0 & 0.125 cu & 0.10 cu \\\hline
 \end{tabular}
     \caption{A summary of parameters varied between three-dimensional runs, namely radius (given in computational units, 1 cu = 0.748 cm), $\alpha$ (given in terms of $\alpha_0 \equiv 4.086\times 10^{-24}$  erg cm${}^{3}$ s${}^{-1}$), and $\beta$, and resolution. Also given are estimates for the cooling length (specifically, the distance required for the temperature to drop by a factor of 10), including all applicable correction factors detailed in section \ref{sec:theory_scales}; isobaric cooling is assumed.
     }
\label{table:3DRuns}}
\end{table}

For the three dimensional runs, we drove two cylindrical jets each with speed 31.58 km s$^{-1}$ from the top and bottom z-boundaries. The jet densities were set to $6\times 10^{16}$ particles per cm$^3$, at a temperature of 720 K, while the ambient medium  density was $1\times 10^{16}$ particles per cm$^3$ at a temperature of 4320 K.  $T_0$ was fixed $2.25\times 10^4$ K across all runs. 
Extrapolated boundary conditions were used in all directions.

Four sets of three-dimensional runs were performed\revision{, summarised in table \ref{table:3DRuns}}.
For the first set of runs we fixed $\alpha$ to  a value
$\alpha_0
\equiv 4.086\times 10^{-24}$  erg cm${}^{3}$ s${}^{-1}$ 
while $\beta$ was varied from $-1.0$ to 3.0 as for  the 1-D case. This corresponds to a cooling time \revision{(as given by equation \ref{eq:tcool})} of $9.5 \times 10^{-6}$ s and a cooling length of 7.5 cm \revision{(more accurate estimates are given in table \ref{table:3DRuns})}.
Inside a cylindrical region of radius 2.0, refinement was allowed to proceed to four levels, with 16 cells per cu (1 cu = 0.748 cm). Refinement was restricted to three levels within radius 3.0 (8 cells per cu) and two levels (4 cells per cu) outside that. The jet radius was set to 2.0 cu, or 32 cells at maximum resolution.

The second set of runs repeated the first set for $\beta = -1.0, +1.0, $ and $+3.0$ using a larger jet radius of 16.0 cu. Refinement was restricted to 3 levels inside the jet radius and 2 levels outside that, so the jet radius was 128 cells at maximum resolution.   \revision{In an effort to suppress numerical instabilities, these runs included the standard diffusion that is part of the 3rd order PPM scheme, which adds dissipation at converging flows.}
The third set of runs repeated the second set of runs with $\alpha=\alpha_h \equiv 10\alpha_0$, giving a cooling time of $9.5 \times 10^{-7}$ s and a cooling length of 0.75 cm. 
The final set of runs consisted of repeating the $\beta=+3.0$ case with $\alpha = 10\alpha_h$ and the $\beta=-1.0$ case with $\alpha=0.1\alpha_0$.

\section{Results} \label{sec:result}

\subsection{One Dimension} \label{sec:results_1D}

We have carried out a series of one dimensional simulations to both verify and extend previous results \citep{stricklandBlondin95}.  In previous work oscillations of radiative shock positions were for $\beta \lesssim 0.75$. We seek to confirm that our models recover these results and explore the consequences of higher values of $\beta$ where the thermal instability may still be active \citep{suzukiVidal15}. 

\begin{figure}
	\includegraphics[width=\columnwidth]{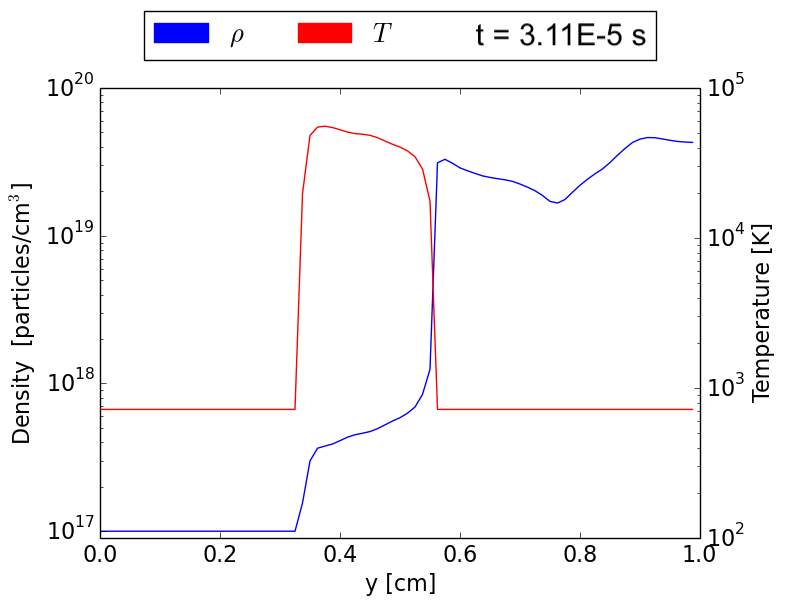}
    \caption{
    The temperature and density vs position after $3.11\times 10^{-5}$ seconds for the $\beta=0$ run. In all runs, the temperature increases rapidly at the front of the shock, which also sees a moderate increase in density. The shocked material cools further from the shock front, before cooling rapidly after some distance. At this point a cold slab of a much higher density begins to build up behind the shock. Some oscillations in the density of this cold slab may form, likely arising from oscillations in the shock itself.
    }
    \label{fig:1DT0}
\end{figure}

We begin by focusing on the structure of radiative shocks as captured in our simulation. Figure \ref{fig:1DT0} shows the run of density and temperature in a radiative shock simulation for a model with $\beta = 0.0$.  This image is taken after $36.84 \hat{t}_\text{cool}$. The temperature is observed to increase at the shock to $T_s = 2.747\times 10^4$ K while density sees the factor of $\sim 4$ as expected for a strong shock. At a distance on the order of one cooling length behind the shock, the temperature returns to its ambient value.  As the temperature (and pressure) drop to its preshock value, the density rises leading to the formation of a "cold slab".  Thus the simulations recovers the classic features expected for a radiative shock.

\begin{figure}
	\includegraphics[width=\columnwidth]{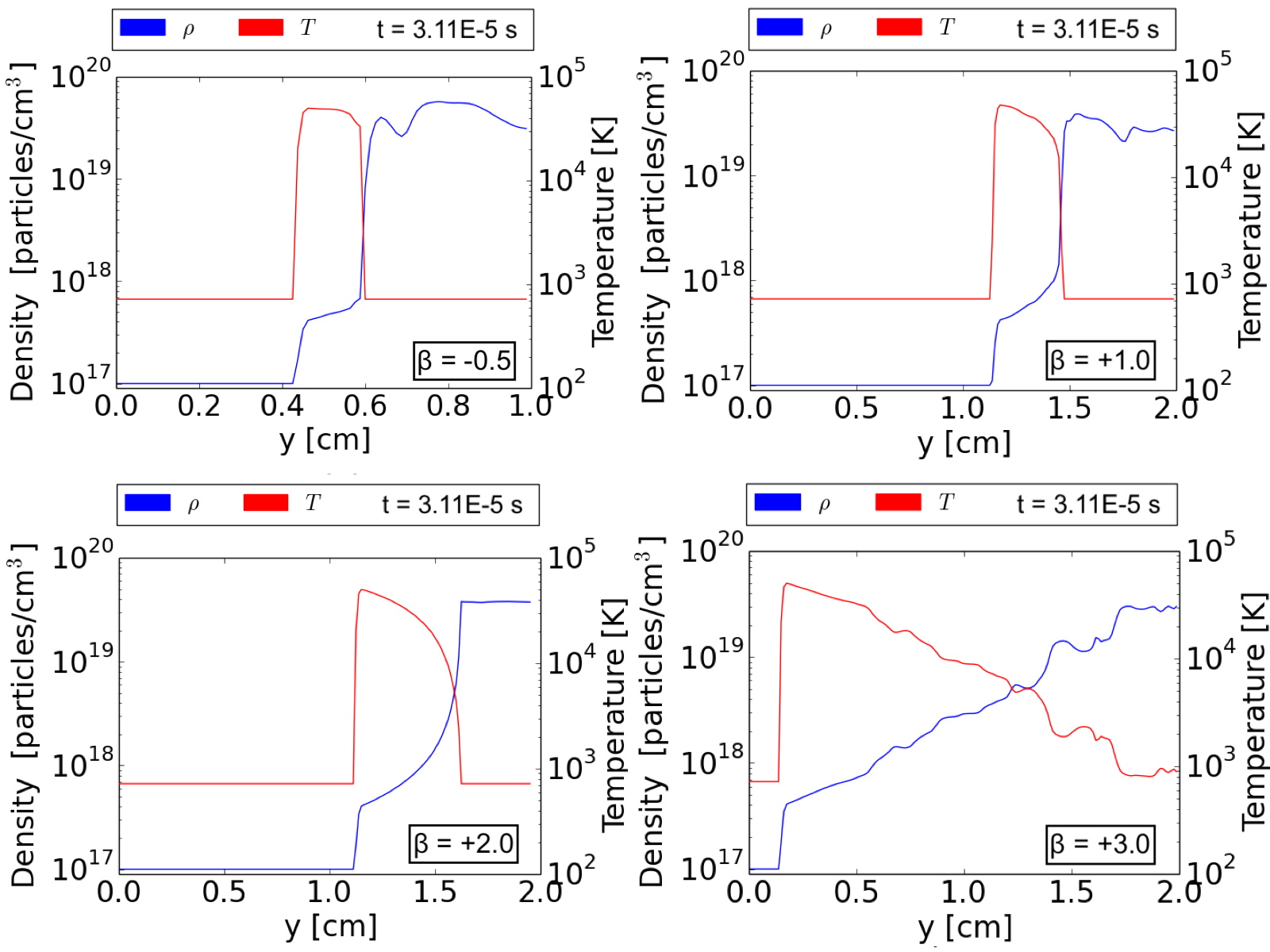}
    \caption{
    The temperature and density vs position after $3.11\times 10^{-5}$ seconds for runs with varying values of $\beta$. As $\beta$ increases, initial temperature decrease is steeper immediately behind the shock, since the cooling rate is higher at larger temperatures. The cold slab however forms at a distance further behind the shock front.
    For $\beta=3$, clumps of matter which are slightly colder and more dense than the surrounding regions begin to form.
    }
    \label{fig:1DTX}
\end{figure}

In Figure \ref{fig:1DTX} we explore the effect of different power law exponents for the cooling.  We show density and temperature distributions for the runs with $\beta = -0.5, +1.0, +2.0, \text{ and } 3.0$. The figure shows that for low $\beta$ \revision{the temperature decrease in the cooling region becomes steeper at distances farther from the shock. This is to be expected: for negative power laws the cooling rate increases 
as the temperature drops, so} a gas parcel passing through the shock will experience a progressively higher  cooling rate
the farther it recedes from the front. For power laws with $\beta > 1$, the cooling rate reduces with distance from the front.

\begin{figure}
	\includegraphics[width=\columnwidth]{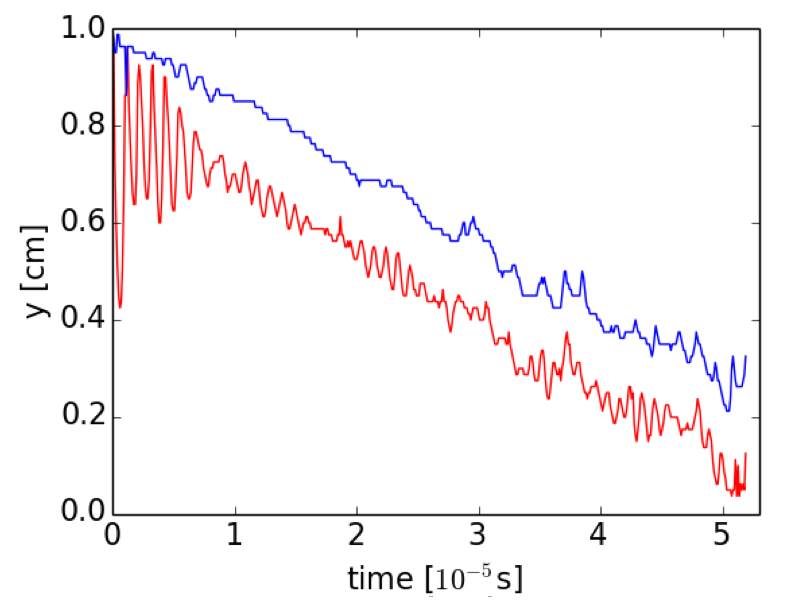}
    \caption{
     A trace of the position of the shock front, measured as the location of peak temperature (red), and the beginning of cold slab (blue) taken for $\beta = 0$. For this cooling law, the position of the shock front has a tendency to oscillate forwards and backwards with an average linear trend \revision{resulting from a buildup of mass in the cold slab}.
    }
    \label{fig:1DP0}
\end{figure}

\begin{figure}
	\includegraphics[width=\columnwidth]{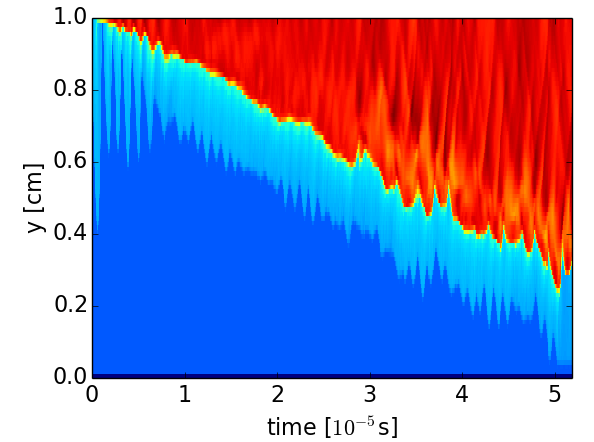}
    \caption{
     A time-space diagram showing density in the $\beta = 0$ case.
    }
    \label{fig:1DST0}
\end{figure}

\begin{figure}
	\includegraphics[width=\columnwidth]{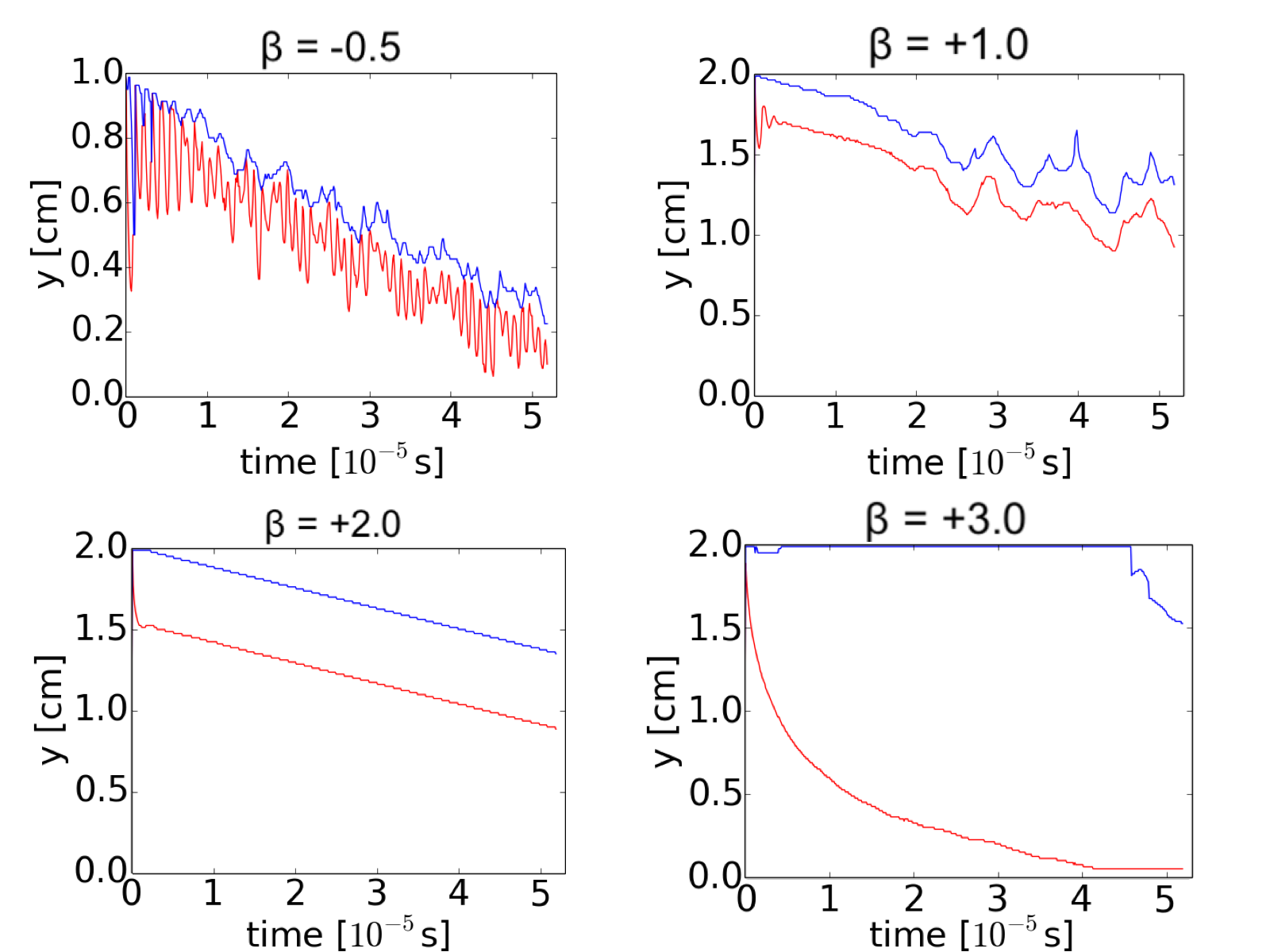}
    \caption{
     Same \typo{as the previous figure but for various} values of $\beta$. As predicted by \citet{stricklandBlondin95}, the oscillations disappear between $\beta = 0.5$ (not shown) and $\beta = 1.0$. At $\beta = 2.0$, the motion becomes nearly perfectly linear. 
    }
    \label{fig:1DPX}
\end{figure}

\begin{figure}
	\includegraphics[width=\columnwidth]{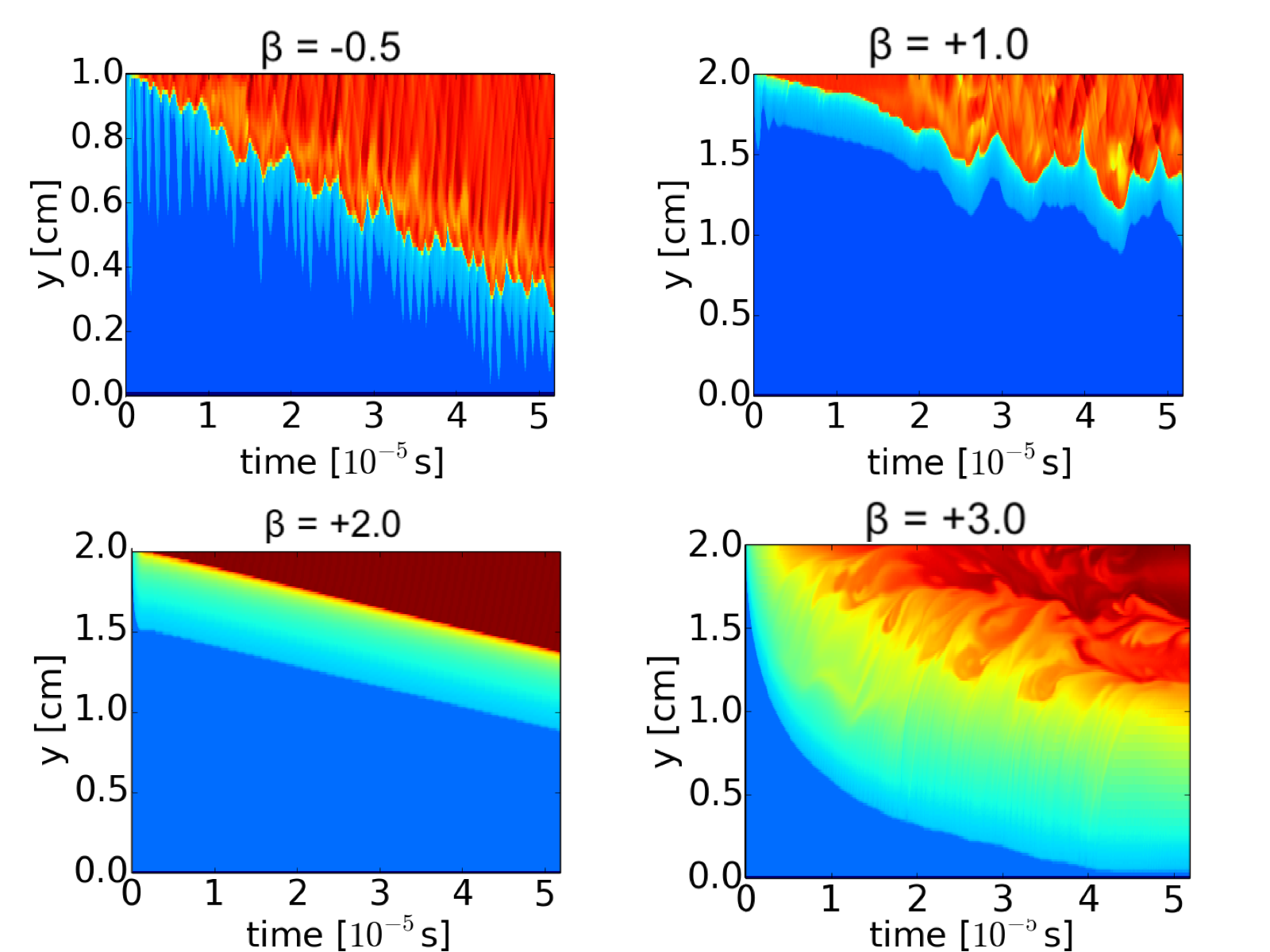}
    \caption{
     A time-space diagram showing density in the cases shown in figure \ref{fig:1DPX}.
    }
    \label{fig:1DSTX}
\end{figure}

The observed size of the cooling region where the temperature drops from $T_s$ to $T_a$ thus deviates from the zeroth order estimate of equation 2 and described in section \ref{sec:theory_scales}.
The increasing size of the cooling region is particularly apparent in the $\beta = 3.0$ case which shows a \typo{log-linear} decrease in $T$ backwards from the shock.  The discrepancy between the estimated post shock temperature used to set $T_0$ and the observed post shock temperature $T_s$ also contributes to the increased cooling length observed for larger values of $\beta$.

Figures \ref{fig:1DT0} and \ref{fig:1DTX} both show how oscillations in the shock front  imprint on the cold slab density.  The variations in $\rho$ for the  $\beta = -0.5, 0.0$ and $1.0$ cases result from oscillations in the shock position. For $\beta\geq 2.0$,  the oscillatory behaviour of the shock front completely vanishes and we see a flat distribution for the cold slab. At $\beta=3.0$, clumps of matter which are slightly colder and denser are observed within the region between the shock front and the cold slab.

\begin{figure}
	\includegraphics[width=\columnwidth]{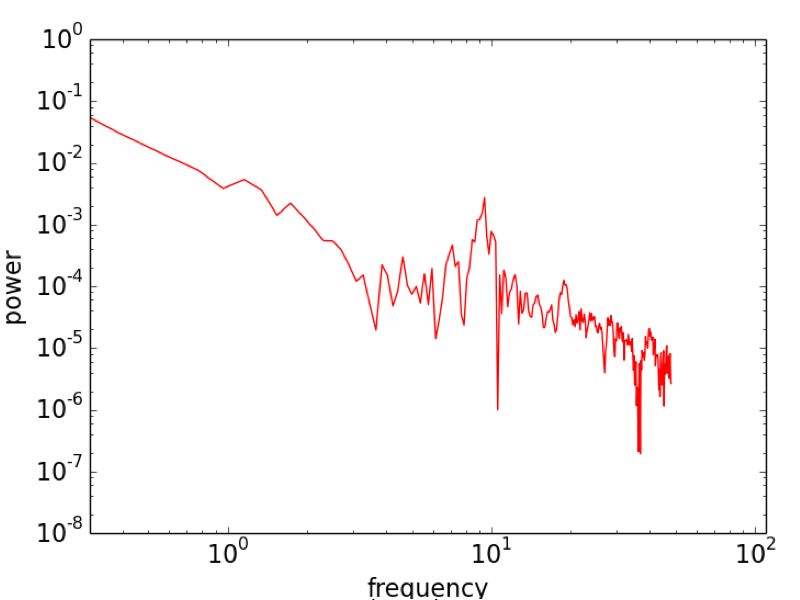}
    \caption{
     Fourier transform of the shock position shown in figure \ref{fig:1DP0}. A strong peak near $\text{frequency} \approx 10$ corresponds to the oscillations mentioned in that figure. 
    }
    \label{fig:1DS0}
\end{figure}

\begin{figure}
	\includegraphics[width=\columnwidth]{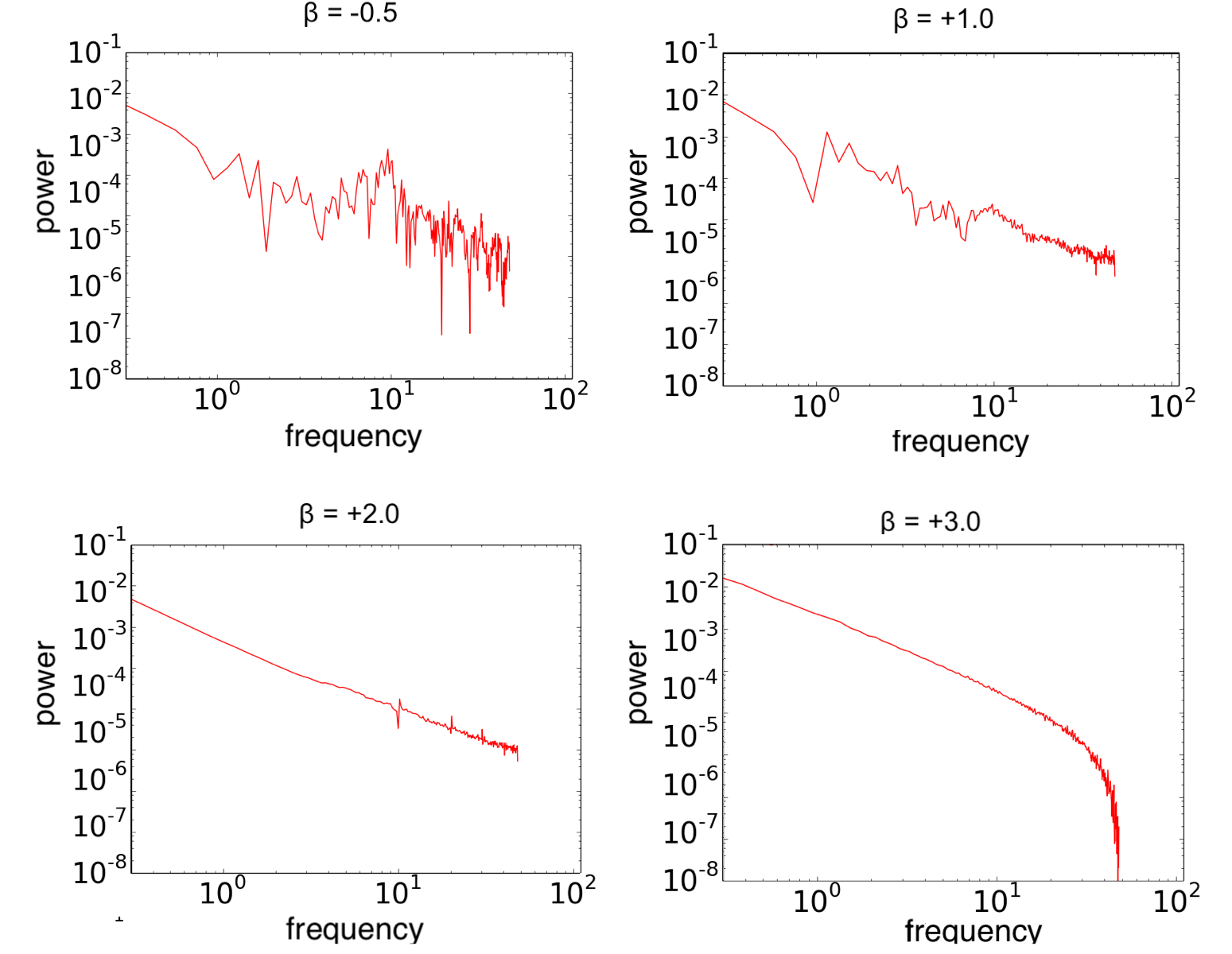}
    \caption{
    Fourier transforms of the shock front positions shown in figure \ref{fig:1DPX}. The strong strong peak near $\text{frequency}\approx 10^1$ is noticeably weaker at $\beta = 1.0$ and vanishes for larger values of $\beta$. \revision{Such a peak} corresponds to the oscillations of the shock front.
    }
    \label{fig:1DSX}
\end{figure}

Figures \ref{fig:1DP0} through \ref{fig:1DSX} show  direct evidence for shock front  oscillations from tracking the shock front position over time. 
Figure \ref{fig:1DP0} shows \revision{the position of the shock front and the cold slab for} the run with $\beta=0.0$. The shock front position was determined by identifying the position of the peak temperature behind the shock in each frame of the simulation. \revision{Meanwhile the position of the cold slab by identifying the point furthest behind the shock front with a temperature above a threshold of 760K, with this threshold being chosen to be slightly higher than the cold slab temperature of 720K in an effort to reduce noise.}
In the $\beta=0.0$ case, strong oscillations in the position of the shock from $t=0$ to $t\sim 0.7$\typo{$\times 10^{-5}$s} are evident, followed by lower amplitude  oscillations 
as the post shock pressure is reduced by cooling and the shock position collapses back towards the cold slab.  \revision{The reduction in amplitude arises from a damping effect provided by the cold slab once it has reached sufficient size.}
The cold slab itself also shows evidence for oscillations.  The entire shock/cold slab structure propagates leftward from the wall with a nearly constant average velocity. \revision{The evolution of structure in the cold slab can be seen in the time-space diagram presented in figure \ref{fig:1DST0}}

We turn now to other values of $\beta$ as shown in \revision{figures \ref{fig:1DPX} and \ref{fig:1DSTX}}.  We find that strong and persistent oscillations persist in both the shock front and cold slab for $\beta <0$.  The oscillations  decay over time for positive $\beta$ and disappear within a few cycles at $\beta=1.0$\revision{, though lower-frequency oscillations driven by cold slab structure return at later times}.  For $\beta=2.0$, no oscillations in the front are seen as the shock moves steadily to the left as the simulation proceeds\revision{, and a constant density is observed throughout the cold slab}. No oscillations are seen in the $\beta=3.0$ simulation either though, as noted above, we do see structure forming in the cooling region.

\revision{Lastly,} we computed the power spectra for the different runs. Figure \ref{fig:1DS0} shows a Fourier transform of the shock front position shown in \ref{fig:1DP0}, with both axes plotted as a natural logarithm. A peak is observed near $\text{frequency}\approx 10$, which corresponds to the oscillations in the shock front. Similarly figure \ref{fig:1DSX} shows a Fourier transform of the shock front positions shown in \ref{fig:1DPX}. The peak corresponding to oscillations \revision{becomes progressively weaker starting $\beta \gtrsim +0.5$, and completely vanishes at $\beta \gtrsim +1.5 $}.

\subsection{Three Dimensions} \label{sec:results_3D}

\begin{figure}
	\includegraphics[width=\columnwidth]{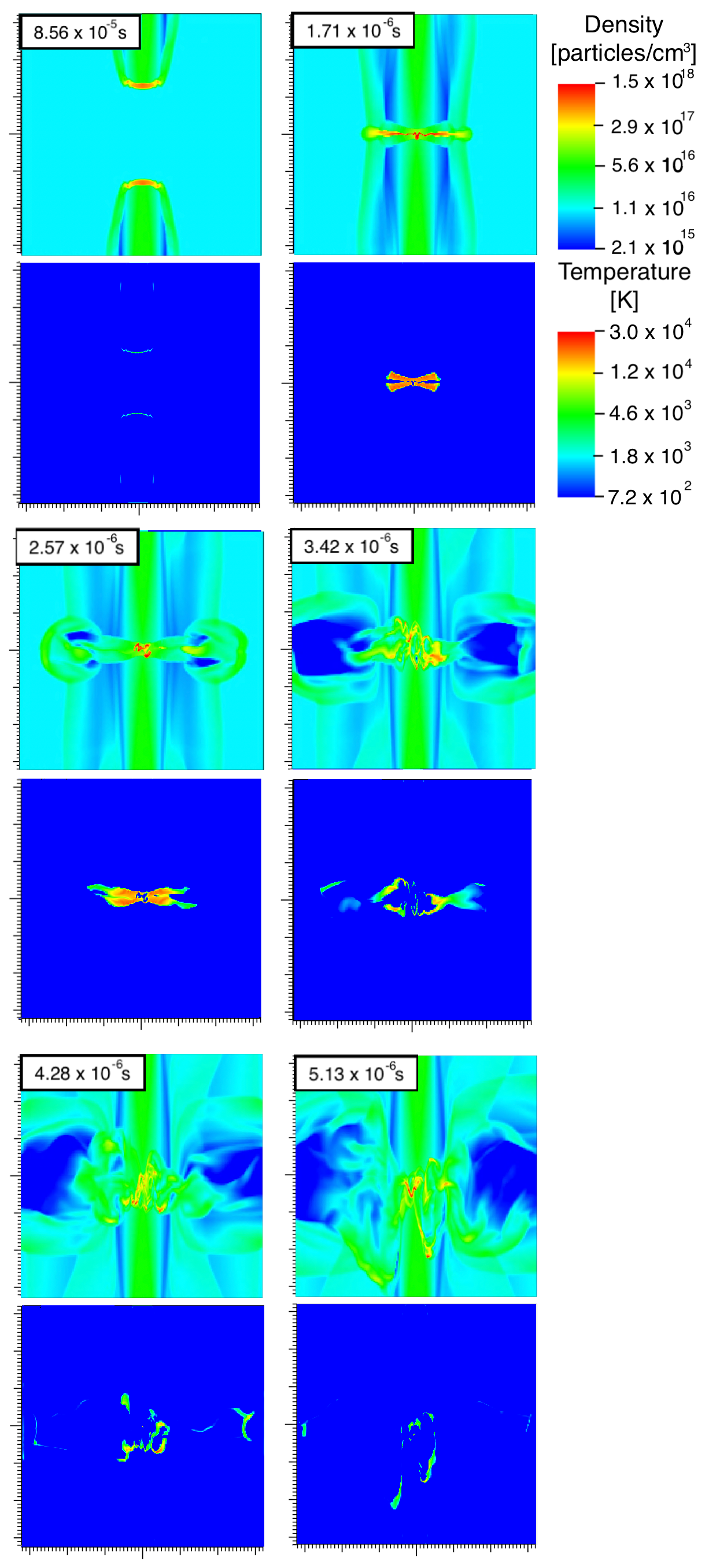}
  \caption{
    \revision{This and all subsequent figures (except as noted) show logarithmic density and temperature midplane slices, with the major tick marks placed at intervals of 10cu = 7.48 cm}. For this figure, the \revision{$\beta=-1.0$} run is shown.
    }
    \label{fig:3DT1}
\end{figure}

To explore the structure and dynamics of the colliding jets, we begin with lower resolution 3-D runs that allow us to track the global evolution of the jet propagation and subsequent collision.  We first examine the case with $\beta=-1.0$ and a jet radius $r = 32 \delta x$  (where $\delta x \sim .047$ cm is the maximum resolution in our AMR grid structure). 

Figure \ref{fig:3DT1} shows the propagation of the jets from their upper and lower injection points. The beam narrows near the heads as a result of the bow shock/cocoon pressure. After the jets collide,  the interaction region forms and material is ejected laterally.  The lateral flow drives its own shocks into the ambient medium which eventually propagate off the grid. Most importantly, the interaction region eventually becomes unstable and  develops strong corrugation-like features which oscillate over time and eventually break into smaller, higher spatial frequency fragments. These structures have a similar morphology to the NTSI instability seen in other studies \citep{Walder98}.

\begin{figure}
	\includegraphics[width=\columnwidth]{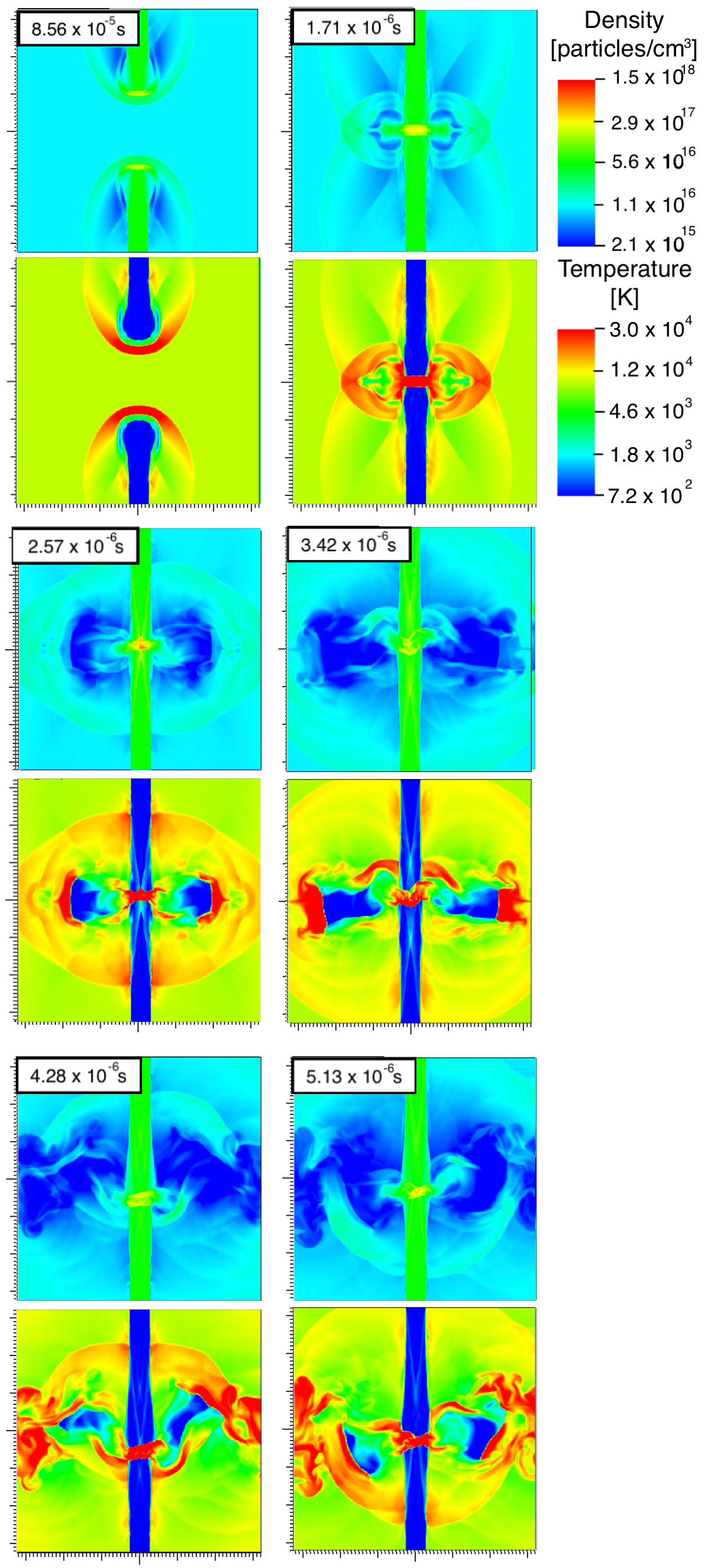}
   \caption{
    Density and temperature slices of the $\beta=3.0$ run.
     In both the initial jet beam as well as the laterally ejected matter, a hot region of shocked gas is followed by a tail of cooled gas which has a lower temperature than the ambient medium.
    }
    \label{fig:3DT3}
\end{figure}

The temperature plots for the $\beta=-1.0$ case show the evolution of the cooling region and cold slab more directly.  Initially, before the collision, we see high temperatures only directly behind the bow shocks.  After the collision, the interaction region shows high temperatures adjacent to the shocks.  Lower temperatures appear along the midplane showing the formation of the cold slab.  Subsequently, 
we see the initiation of a sinusoidal disturbance across the interaction region which then grows to disrupt the region.  By the final frames we see the high frequency corrugations that dominate the interaction region,  showing high temperatures only behind the peaks of the oscillations.

Next we examine the  $\beta=+3.0$, case. Slices from the run are shown in figure \ref{fig:3DT3}. Once again we see bow shocks forming in front of the jets before their collision.  After the collision the formation of the interaction region and the lateral flows are also seen, as in the $\beta=-1.0$ case.  Unlike the negative 
$\beta$
case however, no strong corrugation of the interaction occurs.  Instead the interaction region as a whole appears to oscillate or flicker.  These oscillations leave an imprint in the laterally ejected material which drive a sequence of shocks into the ambient space around the jets.  These shocks are particularly apparent in the temperature plots.  In the temperature plots  we can see the formation of a cold slab, but the slab material  does not reach the ambient temperature.  This is \typo{because the} positive slope of the cooling curve weakens the cooling rate as the temperatures behind the shocks fall (equation \ref{eq:correction}). We note also that the ambient medium does not cool much below its original temperature for the same reason.

\begin{figure}
	\includegraphics[width=\columnwidth]{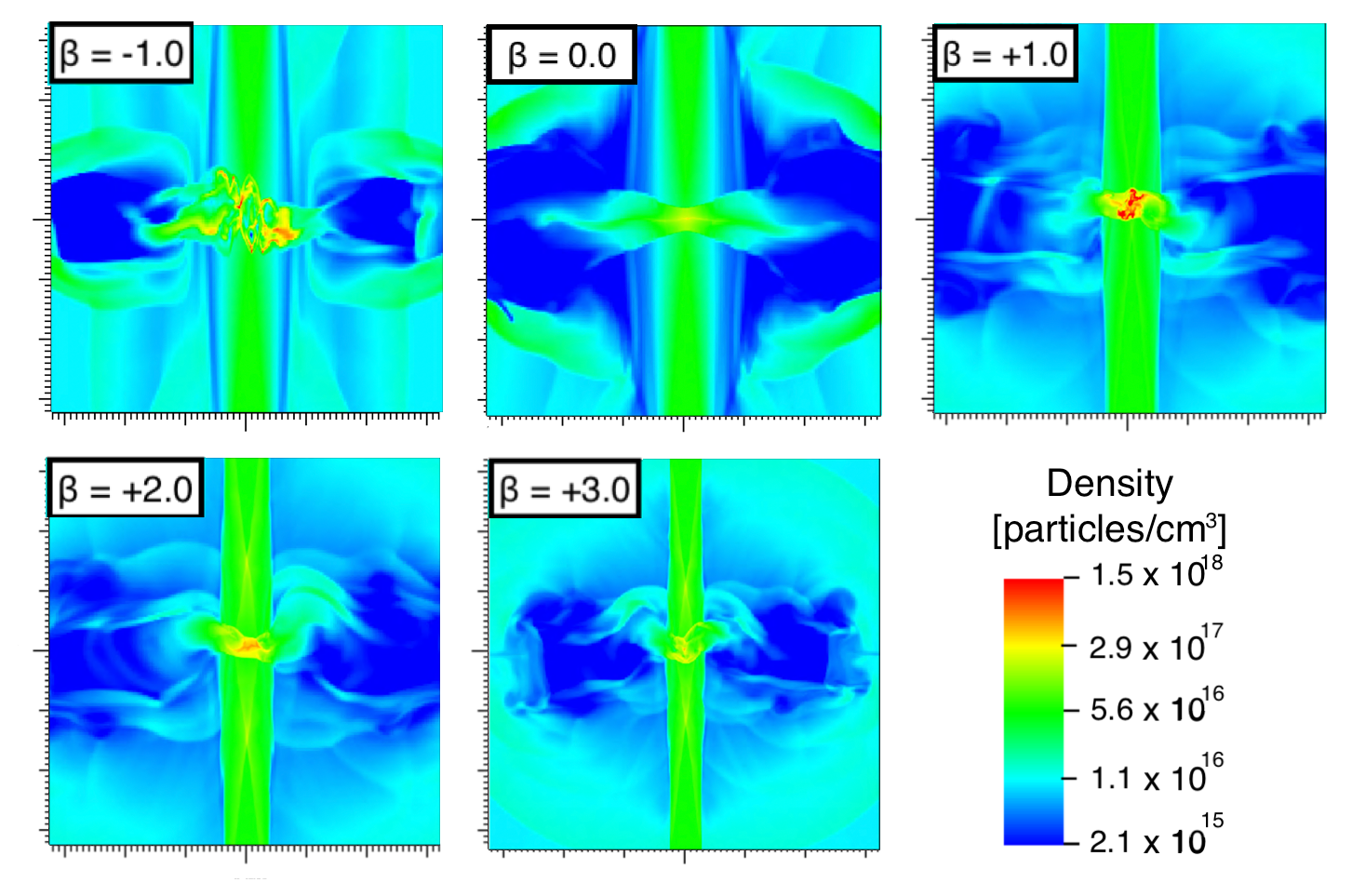}
    \caption{
    Density slices for runs with integer values of $\beta$ after $3.42\times 10^{-5}$ s.
    The interaction region is observed unstable for $\beta=-1.0$, in which the interaction region breaks apart into several smaller fragments.
    }
    \label{fig:3DS2}
\end{figure}

In figure \ref{fig:3DS2} we show a larger set of runs with different values of $\beta$. As $\beta$ increases from $-1$ to $3$ we see the nature of the interaction change.  As we have already explored, for $\beta = -1.0$ the interaction region is disrupted.  For $\beta = 0.0$, or constant cooling, such disruption is no longer present however. As $\beta$ increases further we do see the interaction region oscillate as a whole or flicker.  Because the interaction region behaves as a more cohesive unit for $\beta \gtrsim 1.0$, 
more coherent lateral ejections appear, and form  
a sequence of shocks. These are most
conspicuous  in the temperature plots, as in Figure \ref{fig:3DT3}).
Finally, we note that because the 3-D simulations discussed so far were designed to also capture the lateral flows, they have a lower resolution.  We next turn to a set of simulations with higher resolution across the jet to study the nature of its instabilities.

We now explore runs with a jet radius of $r = 128 \delta x$  (where $\delta x \sim .093$ cm is the maximum resolution for these runs).  Thus \typo{these runs have 4 times higher resolution} that those just discussed. For each value of $\beta=-1.0$ we made two runs,  with  low and high values of cooling strength coefficient $\alpha$ respectively.  For our low value we use $\alpha_0$ and for our high value of $\alpha_{h} = 10 \alpha_0$.

\begin{figure}
	\includegraphics[width=\columnwidth]{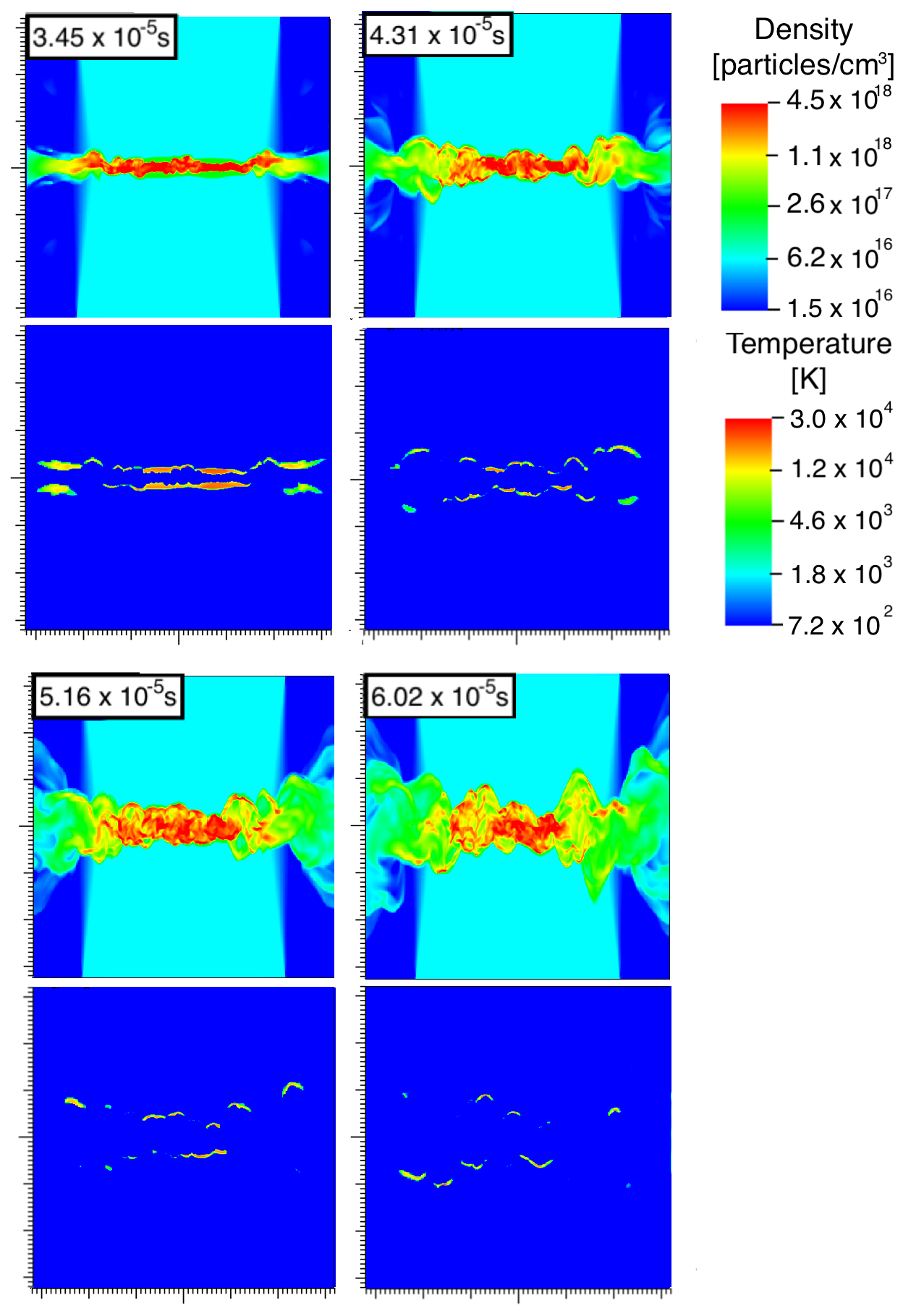}
    \caption{Density and temperature midplane slices for the $\alpha=\alpha_0, \beta=-1.0$ run with increased jet radius.
    Note that gas outside of the cooling region approaches the isothermal limit since cooling strength increases at lower temperatures}
    \label{fig:3DL1}
\end{figure}
\begin{figure}
	\includegraphics[width=\columnwidth]{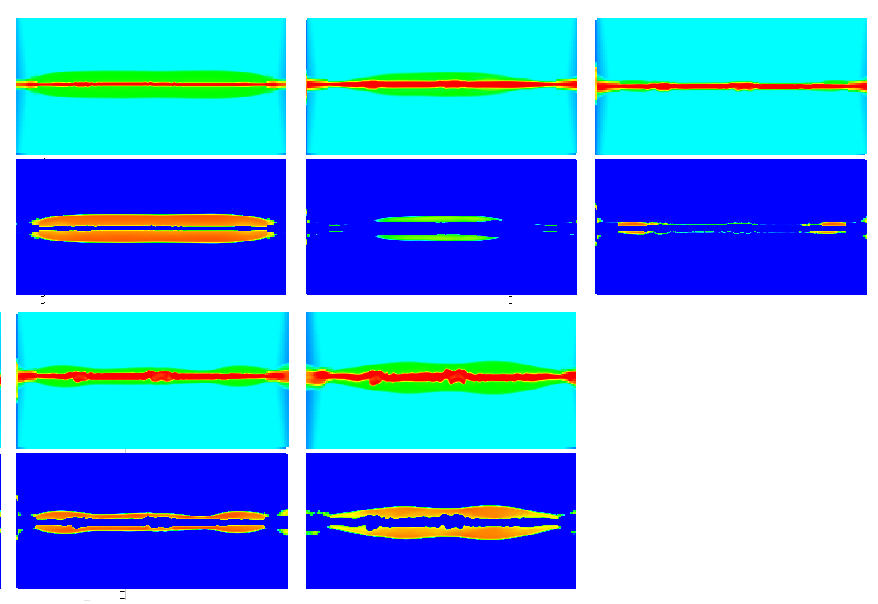}
    \caption{
    Oscillations in the shock front position  as seen in the 3D run with $\alpha=\alpha_0, \beta=-1.0$, and larger jet radius.
    }
    \label{fig:3DOsc}
\end{figure}
\begin{figure}
	\includegraphics[width=\columnwidth]{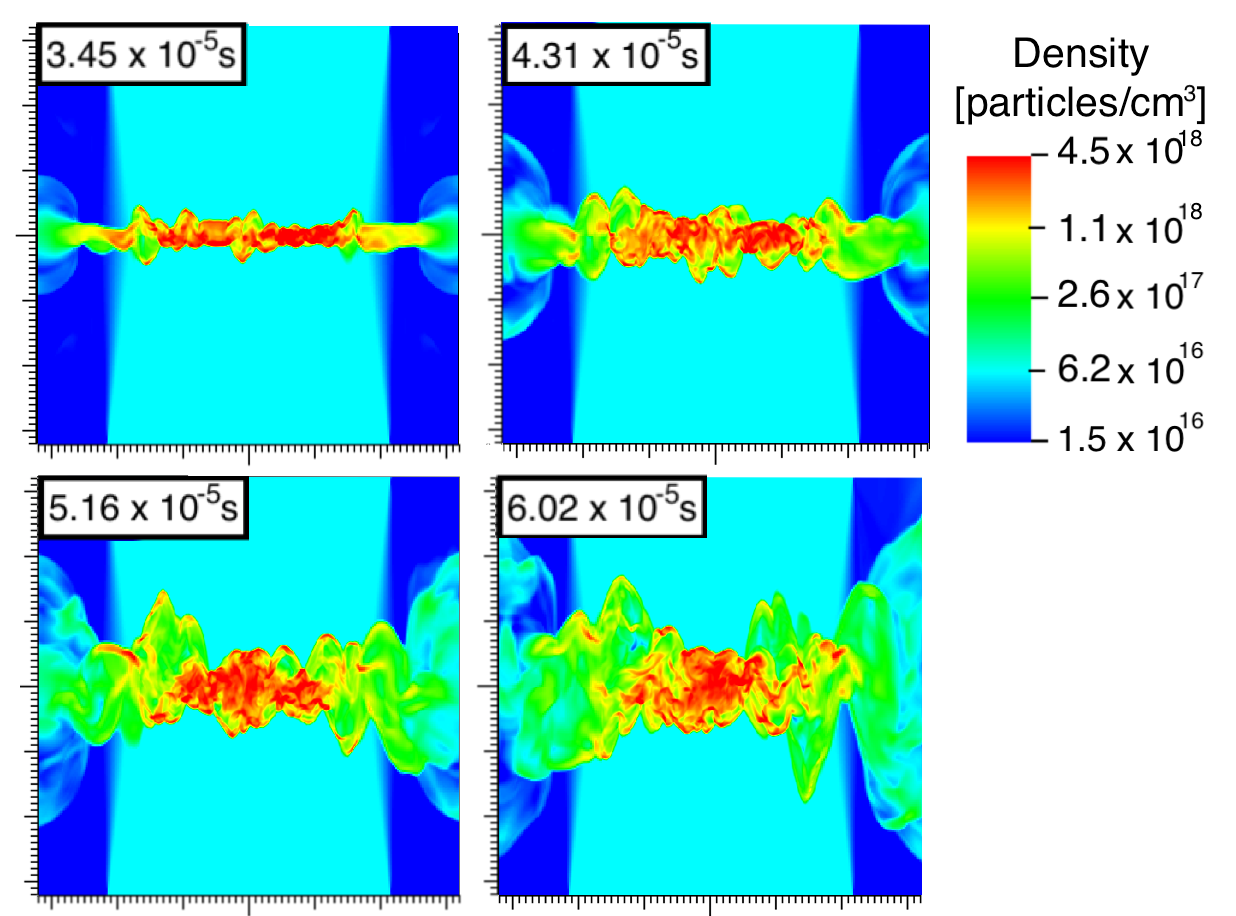}
    \caption{Density slices for the $\alpha=10\alpha_0, \beta=-1.0$ run with increased jet radius. Temperature is omitted since the cooling length is comparable the grid resolution.}
    \label{fig:3DLC1}
\end{figure}
\begin{figure}
	\includegraphics[width=\columnwidth]{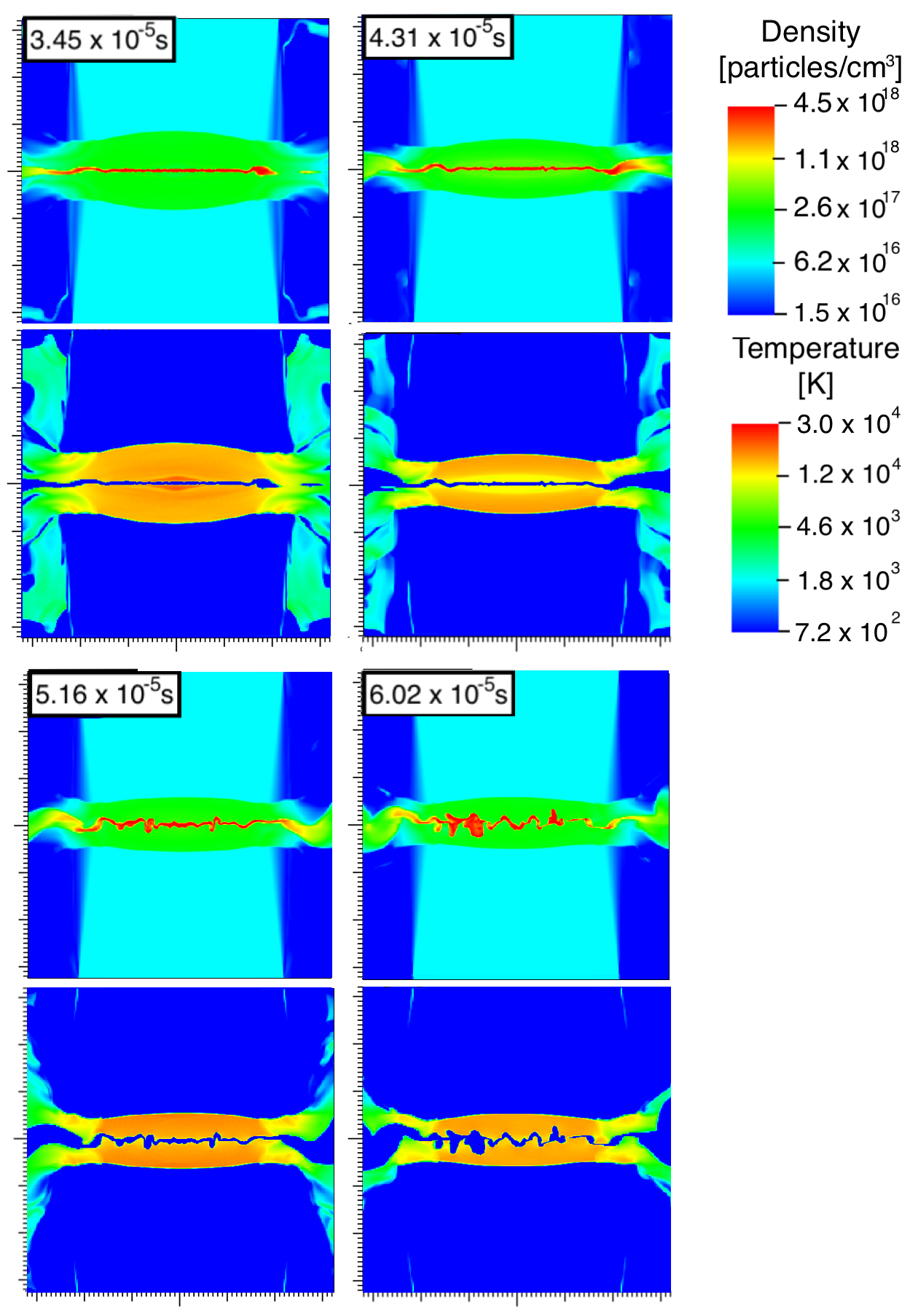}
    \caption{Density and temperature slices for the $\alpha=0.1\alpha_0, \beta=-1.0$ run with increased jet radius. }
    \label{fig:3DLH1}
\end{figure}

We first discuss the $\beta=-1.0$ low $\alpha$ case  shown in figure \ref{fig:3DL1}. The larger jet radius and higher resolution allows the internal structure of the interaction zone to be more readily observable. We now  see  the formation of the cold dense slab (in red on the density maps) behind the two shocks which define the limits of the interaction region.  We also see the effect of the radiative shock instability as the cooling distance between the shock and the cold dense slab collapses leaving the shock in close proximity to the cold dense slab.  This collapse then triggers the NTSI as the shocks begin to show the growth of bending-mode perturbations.

In figure \ref{fig:3DOsc} we show a series of more closely spaced frames in time with a spatial zoom in on the interaction region.  This figure allows us to temporally resolve the oscillations of the shock fronts\revision{, which show a period of $8.6\times 10^{-6}$s}.  Note in particular, how the oscillation of $d_\text{cool}$ separating the cold slab from the shock in different parts of the interaction region are out of phase. Changes in $d_\text{cool}$ at centre of the jet lag behind the oscillations at the edge of the jet by about a quarter period.

When we rerun the simulation with the higher value of $\alpha$ (shown in figure \ref{fig:3DLC1}) the cooling becomes stronger and $d_\text{cool}$ decreases.  We once again see the onset of the NTSI as the shock collapses on to the cold slab. In this case however, the size of the cooling region is reduced to a scale smaller than the grid resolution so we can no longer follow the oscillations of the radiative shock instability.

 When we instead rerun the simulation with $\alpha=0.1\alpha_0$ (figure \ref{fig:3DLH1}) the oscillation slows such that a single oscillation period exceeds the duration of the simulation, so NTSI (if present) does not begin until after the duration of the run. It is worth noting however that the cold slab in this case develops structure rich in bending modes, likely imposed from the early jet shaping, making NTSI likely to occur once the cooling region collapses.

\begin{figure}
	\includegraphics[width=\columnwidth]{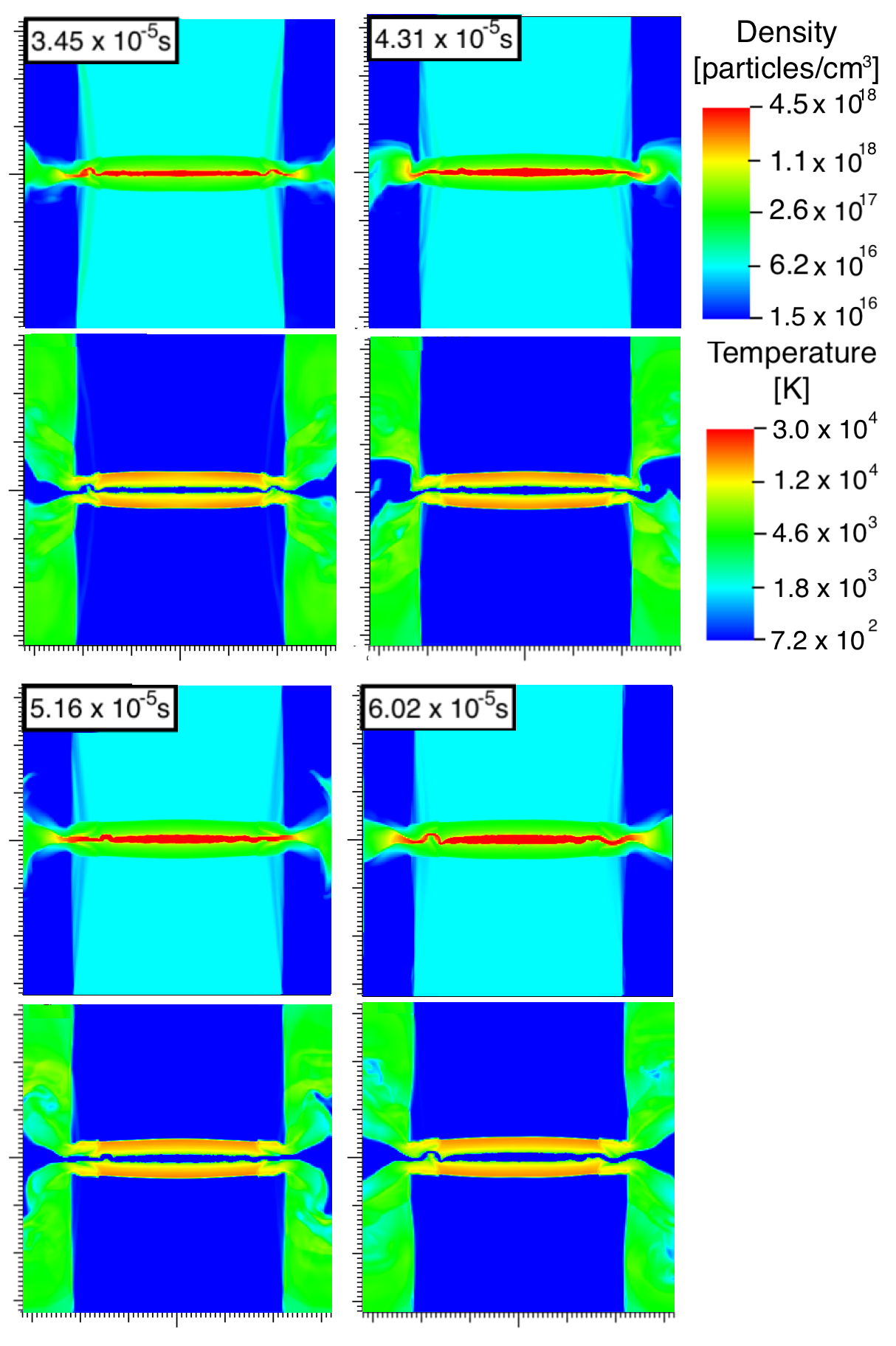}
    \caption{
    Density and temperature slices  for the $\alpha=\alpha_0, \beta=+1.0$ run with increased jet radius.
    }
    \label{fig:3DL2}
\end{figure}
\begin{figure}
	\includegraphics[width=\columnwidth]{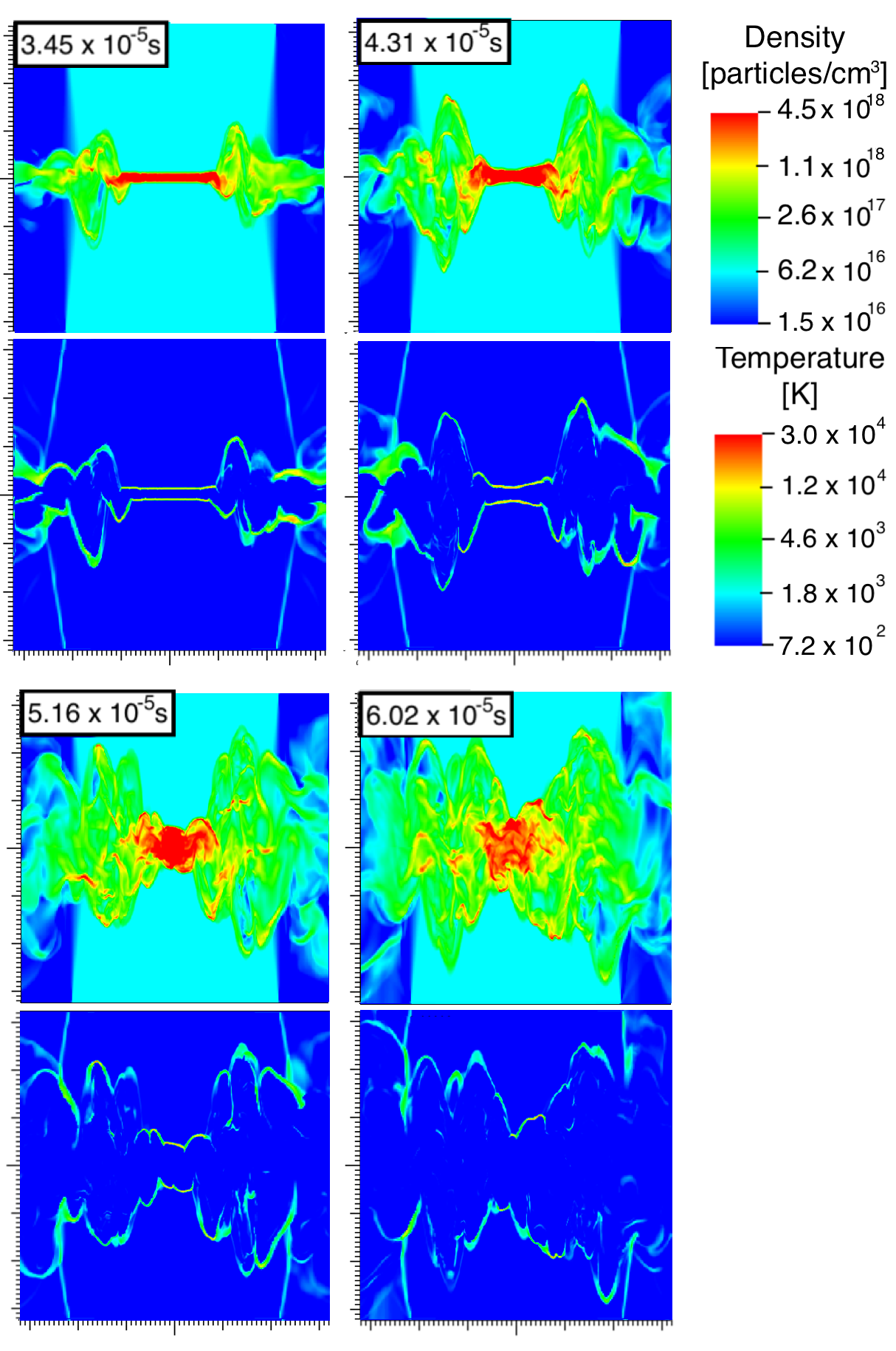}
    \caption{
    Density and temperature slices  for the $\alpha=10\alpha_0, \beta=+1.0$ run with increased jet radius.
    }
    \label{fig:3DLC2}
\end{figure}

We now turn to the $\beta = +1.0$ case which exhibits pronounced differences as $\alpha$ changes. For low  $\alpha$ (shown in figure \ref{fig:3DL2}) the separation between the shocks and the cold slab remains initially stable.  This is expected as $\beta = +1.0$ is too large to initiate the radiative shock instability.  Since $d_\text{cool}$ is large enough to ensure no contact between the shocks and the cold slab, the global bending modes of the NTSI are not triggered.  However, as the simulation progresses we do see ripples growing along its length of the cold slab with a coherent wavelength comparable to its thickness $\lambda \sim d_\text{slab}$.  It should be noted that if the simulation is run without diffusion, a large perturbation develops along the axis of the jet, likely seeded by grid effects.

\begin{figure}
	\includegraphics[width=\columnwidth]{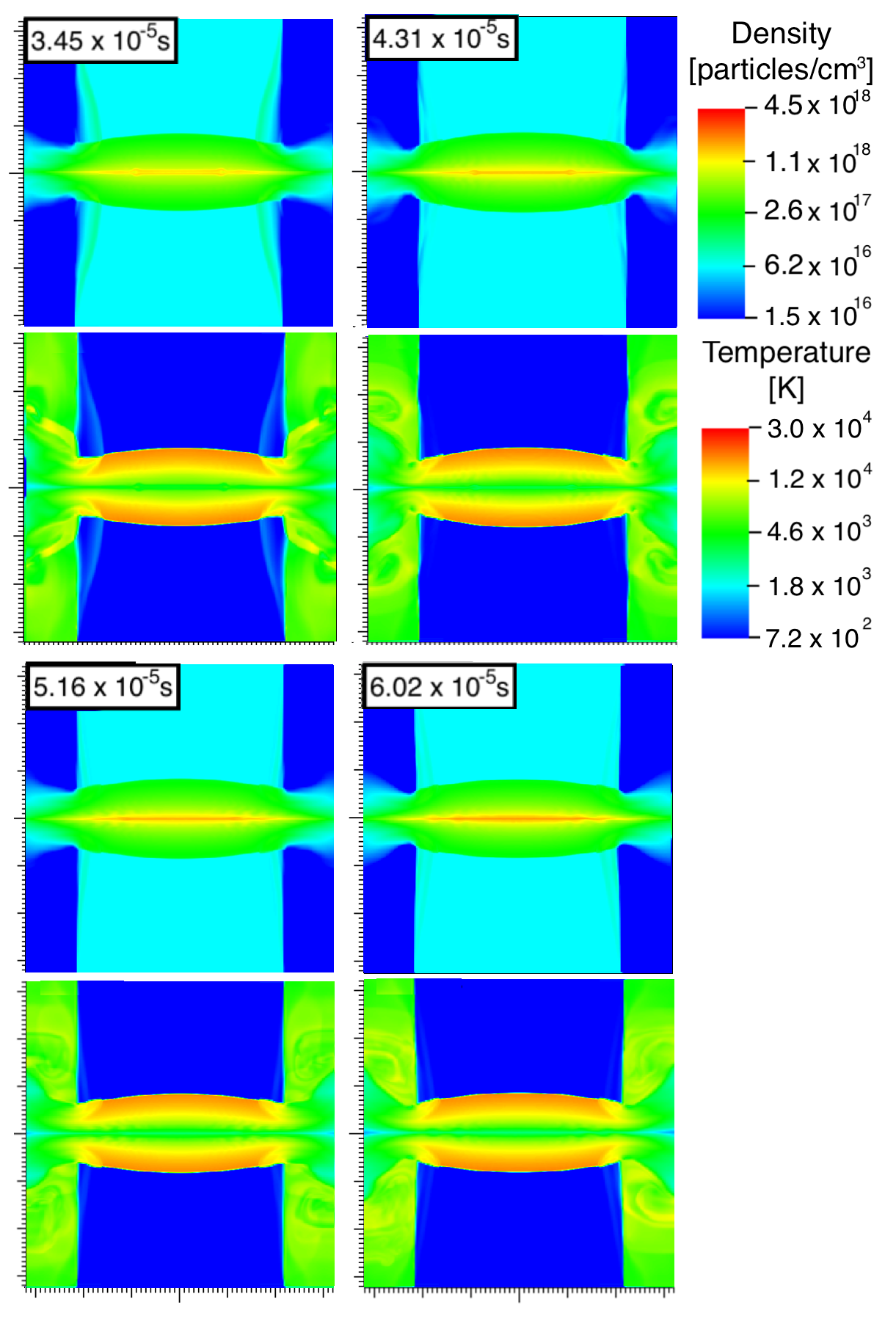}
    \caption{
    Density and temperature slices for the $\alpha=\alpha_0, \beta=+3.0$ run with increased jet radius.
    }
    \label{fig:3DL3}
\end{figure}
\begin{figure}
	\includegraphics[width=\columnwidth]{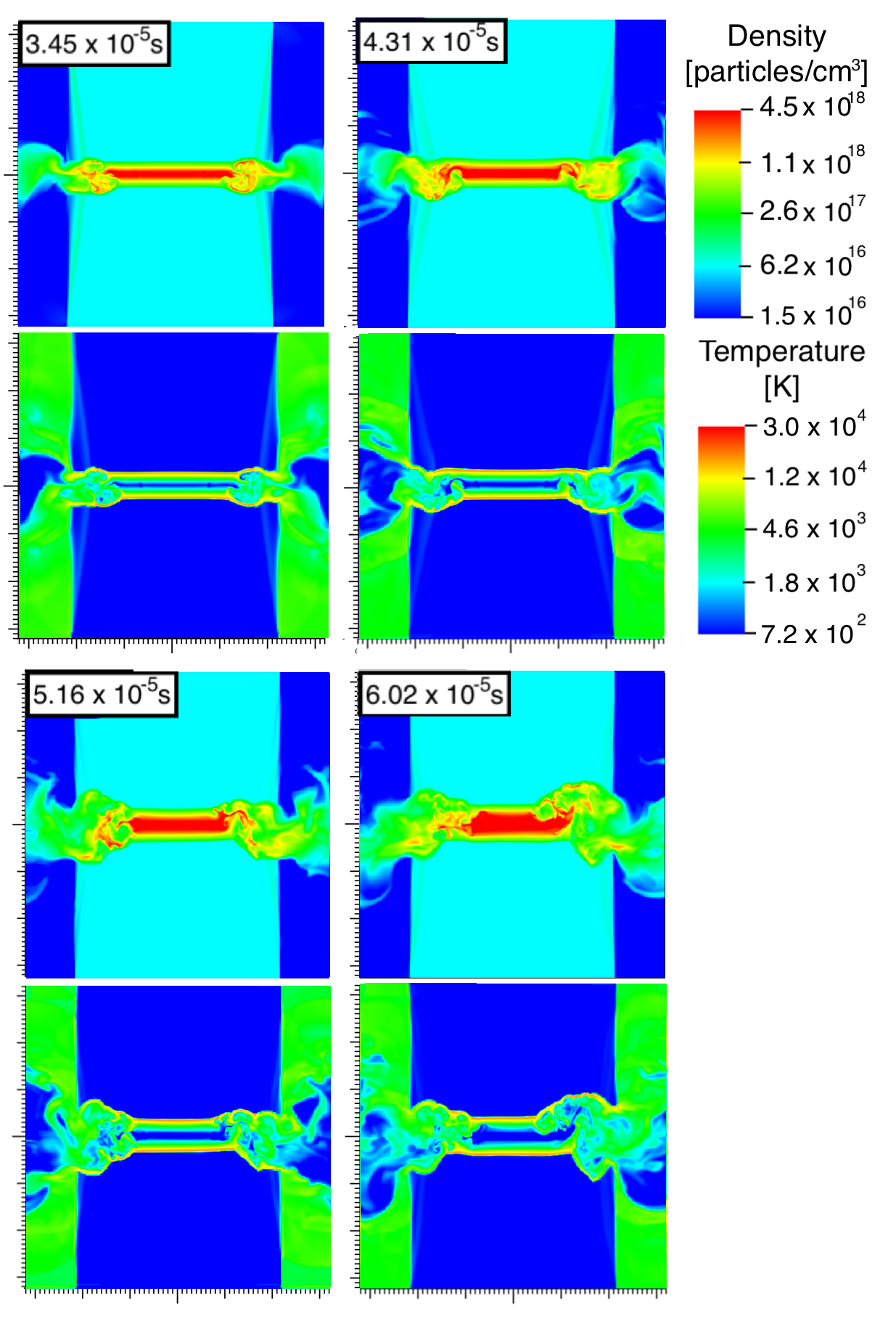}
    \caption{
    Density and temperature slices for the $\alpha=10\alpha_0, \beta=+3.0$ run with increased jet radius.
    }
    \label{fig:3DLC3}
\end{figure}
\begin{figure}
	\includegraphics[width=\columnwidth]{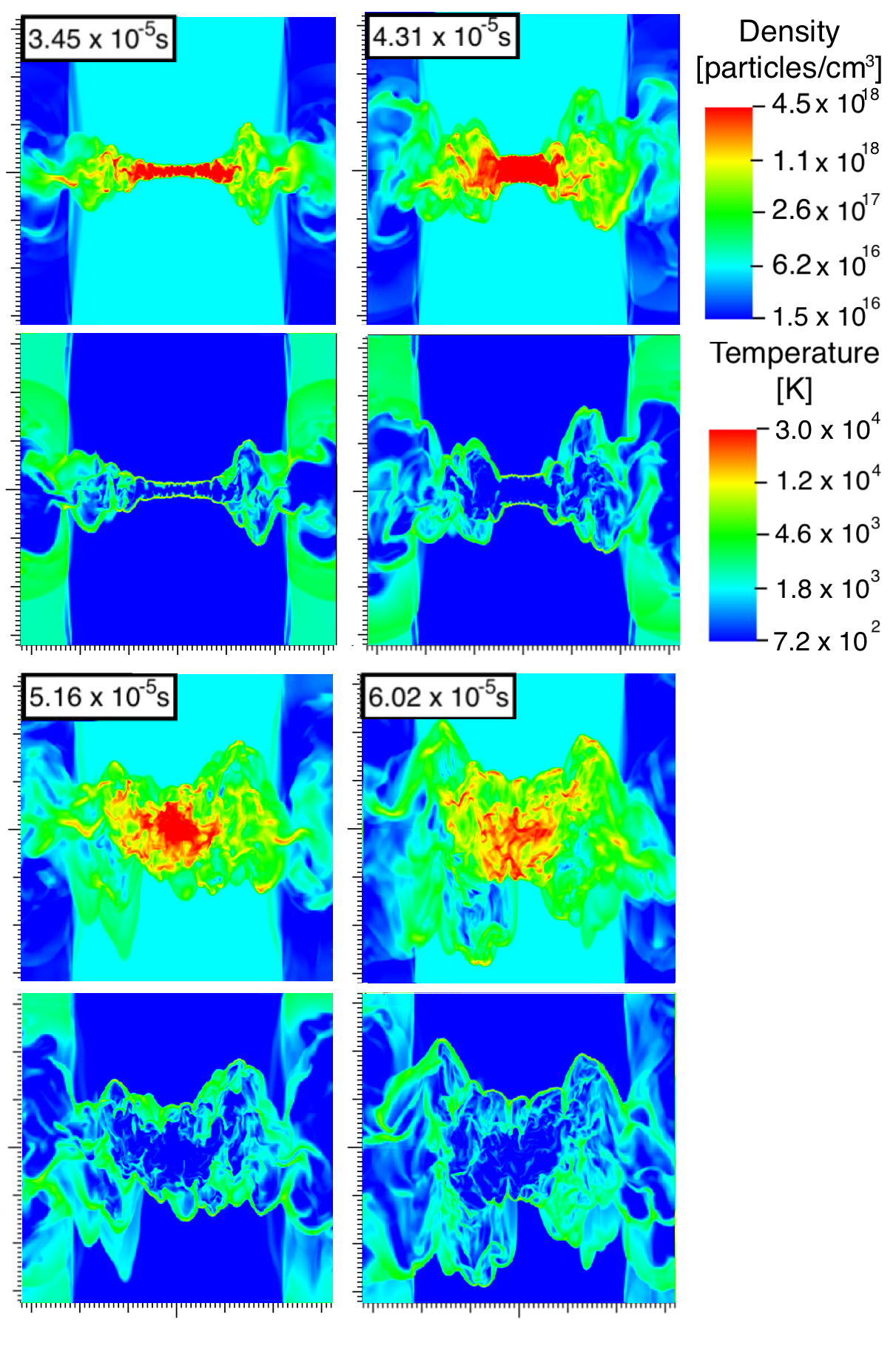}
    \caption{
    Density and temperature slices for the $\alpha=100\alpha_0, \beta=+3.0$ run with increased jet radius.
    }
    \label{fig:3DLU3}
\end{figure}

In figure \ref{fig:3DLC2} we show the $\beta = +1.0$ simulation run with the larger  $\alpha=\alpha_h$.  In this case the lower value of $d_\text{cool}$ means the shock discontinuity is closer to the cold slab.  This implies the shock geometry \typo{will respond} to any variations of the cold slab and once again we see the NTSI is triggered. Growth of the bending modes begins at large radii from the jet axis where the jet head was imprinted with features 
from its propagation before the collision. The NTSI appears to begin at these features and propagate inwards. Such behaviour was also seen for the $\beta=-1.0$ case but was less pronounced there, likely due to the NTSI being triggered from radiative shock oscillations along the entire length of the cold slab. 

Finally we examine the $\beta=+3.0$ case. The $\alpha=\alpha_0$ (figure \ref{fig:3DL3})  case does not exhibit instabilities since the cooling length increases at higher $\beta$. The $\alpha= \alpha_h$ (figure \ref{fig:3DLC3}) case remains stable near the centre, while jet shaping seeds instabilities near the edges which progress slightly inward at later times. 
Since the cooling length increases at higher $\beta$, the separation between the shock and the cold slab is larger than what is seen in the $\beta=+1.0$ case, weakening the effects of NTSI and slowing the rate at which instabilities grow and progress inward.
When $\alpha$ is increased to $10\alpha_h$ (figure \ref{fig:3DLU3}), the cooling length is once again small enough to fully allow for NTSI in a manner similar to the $\beta=+1.0, \alpha=\alpha_h$ case.
 It should be noted that multiple instabilities were seen when the $\beta=+3.0$ case was run without diffusion: carbuncles were seen to form behind the shock discontinuity, perturbations grow in the cold slab, and at later times local curvature distortions arise in the cooling region. 
 
 We also ran a case with an even higher value of $\alpha$ which is \revision{shown in  figure \ref{fig:3DLU3}}. we find as expected that the NTSI is triggered early as the shock collapses on to the cold slab as occurred in the strong cooling case for $\beta = +1$.

\section{Interpretation with respect to theoretical instabilities} \label{sec:discuss}

The purpose of this paper was to initiate a set of studies relevant to the experiments of \citep{suzukiVidal15} which themselves addressed issues associated with the evolution of bow shocks in astrophysical jets.  In particular, bow shocks occurring in environments dense enough to allow for strong radiative cooling are often observed to show fragmentation and clumpiness \citep{Hartigan11}.  Such fragmentation was also seen in the experiments of \citep{suzukiVidal15} who attributed break-up of the bow shock to thermal instabilities.  In this paper, 
we have focused solely on  the nature of instabilities in colliding flows abstracting the problem away from either the laboratory or astrophysical situation by using a simplified power-law cooling curve and using identical flows without magnetic fields.  In what follows we discuss the extent to which the 
different unstable modes introduced in section \ref{sec:theory_insta} are evident in our simulations.

We begin with  the thermal instability \citet{field65}.
While \citet{suzukiVidal15} \typo{argued the thermal mode} as the source of their shock structure, we do not see evidence of this instability  in our simulations. The thermal mode should occur for $\beta < +2.0$. In our  one-dimensional runs (figures \ref{fig:1DT0} and \ref{fig:1DTX}) we do not see strong density "clumping" in either the cooling region or the cold slab.  The density variations that are seen do not grow significantly and are not accompanied by temperature changes and are thus not isobaric. These variations are likely the result of the motions of the shock front leaving an imprint on the cold slab. 

Next,  we turn to the radiative shock instability first described by \citet{Langer81}. Our one-dimensional simulations instability limits ($\beta < 1$ for the instability to grow) agree with the results found by \citet{stricklandBlondin95}. The presence of oscillations can be seen in figures \ref{fig:1DP0} and \ref{fig:1DPX}, where the location of the shock front is seen to oscillate for those runs for which the value of $\beta$  is below the critical value.  

Further, while variations appear in both the location of shock front and the cold slab with time, the oscillations are much more pronounced for the shock front. This is consistent with the oscillations being driven by imbalances of pressure at the shock front driving oscillations in the location of the shock, while the cold slab is in approximate pressure equilibrium with the cooling region and thus its variations are likely a residual effect of the shock front oscillations.

Most importantly we are able to observe these oscillations in the shock fronts in the 3D runs for low $\beta$. Observation of the radiative shock instabilities oscillations require resolution of the cooling region and hence depend on the strength of cooling which was controlled in our simulation by the parameter $\alpha$. Neither the $\beta=1.0$ or $\beta=3.0$ runs showed strong shock oscillations as expected based on the stability limits confirmed in the 1-D simulations. 

Of particular significance to our conclusions is the presence of bending modes in 3D runs, consistent with the NTSI. We note that this  instability is not directly dependent on $\beta$ however the strength of cooling does determine the structure of the interaction region and so allows the "thin shell" conditions in the NTSI to appear. What matters for triggering the NTSI is that cooling length $d_\text{cool}$ is small enough that perturbations in the cold slab are communicated to the shock waves bounding the interaction regions.

To explicate this point, note that in our lower cooling strength ($\alpha_0$) cases the NTSI is only observed in runs when the value of  $\beta$ is below the critical value given by \citet{stricklandBlondin95}. In these cases even if the nominal $d_\text{cool}$ is large, the shock oscillations mean there are periods where the cooling region collapses and the NTSI can be triggered. 
However if $\alpha$ is lowered even further the frequency of shock oscillations decreases, resulting in NTSI not being triggered until a much later time. Thus the $\beta<\beta_\text{cr}$ and $\beta>\beta_\text{cr}$ recover similar behaviour at finite times \typo{as the} in the as cooling approaches the adiabatic limit.

For $\beta=+1.0$ and weaker cooling (figure \ref{fig:3DL2}) $d_\text{cool}$ is large and the NTSI does not occur. The distance between the shock and the cold slab remains constant, and even though perturbations do eventually appear in just the cold slab,  the entire interaction region does not \typo{participate in them}.  Only for the $\alpha_h$ case (figure \ref{fig:3DLC2}) do we see global disruption of the interaction by the NTSI.  The same holds true for the $\beta=+3.0$ case (figures \ref{fig:3DL3},\ref{fig:3DLC3}, \ref{fig:3DLU3}), except that an even higher value of $\alpha$ is required to fully exhibit the effects of NTSI. Note that while all runs with the same value of $\alpha$ have the same cooling length estimate given by equations \ref{eq:dcool} and \ref{eq:tcool}, the correction factor given by equation \ref{eq:correction} increases for higher values of $\beta$.

Finally we note that the NTSI requires an initial perturbation  for growth to occur. For  $\beta<\beta_\text{cr}$ it is likely that the oscillations from the radiative shock instability seed an initial perturbation. While the radiative shock instability itself only provides time variation, a combination of edge effects and bow shock curvature results in oscillations in the middle of the jet, lagging about a quarter-period behind the oscillations in the edge (see figure \ref{fig:3DOsc}). This produces a spatial variation which  would allow  the non-linear effects to begin. The fundamental mode dominates this initial variation, but  shorter wavelengths dominate growth (see figures \ref{fig:3DL1}).  This is consistent with the instability having a faster growth rate for shorter wavelengths (equation \ref{eq:NTSI}), thus preferentially amplifying these modes relative to the fundamental. 

For the bending modes observed in cases with $\beta>\beta_\text{cr}$, strong cooling runs may be seeded by early jet shaping (figure \ref{fig:3DSchematic}). As perturbations grow in the outer regions, they cause disturbances in progressively inward regions, consistent with the behaviour observed in figure \ref{fig:3DLC2}. The detailed effects of curvature and jet shaping on shock structure will be examined in a future paper.

We conclude that the strong fragmentation of the interaction region in colliding flows leading to strong clumping is most easily promoted by a combination of the radiative shock instability and the NTSI.  If the cooling is strong enough and local perturbations are present however, the NTSI can be \typo{triggered by strong} modifications in the interaction region on its own. Because cooling curves such as those used for both astrophysical and laboratory studies show a range of slopes, $\frac{\operatorname{d}\Lambda(T)}{\operatorname{d}T}$, both in terms of sign and \typo{magnitude}, future studies should attempt to isolate the temperature and strength (i.e. $\alpha$) regions where conditions lead to the radiative shock instability and NTSI and drive fragmentation.

\section{Acknowledgments}

This work used the computational and visualization resources in the Center for Integrated Research Computing (CIRC) at the University of Rochester. Financial support for this project was provided by the Department of Energy grant GR523126, the National Science Foundation grant GR506177, and the Space Telescope Science Institute grant GR528562.

\section{Data Availability}

The data underlying this article will be shared on reasonable request to the corresponding author.

\bibliography{main.bib}

\begin{thebibliography}{}
\makeatletter
\relax
\def\mn@urlcharsother{\let\do\@makeother \do\$\do\&\do\#\do\^\do\_\do\%\do\~}
\def\mn@doi{\begingroup\mn@urlcharsother \@ifnextchar [ {\mn@doi@}
  {\mn@doi@[]}}
\def\mn@doi@[#1]#2{\def\@tempa{#1}\ifx\@tempa\@empty \href
  {http://dx.doi.org/#2} {doi:#2}\else \href {http://dx.doi.org/#2} {#1}\fi
  \endgroup}
\def\mn@eprint#1#2{\mn@eprint@#1:#2::\@nil}
\def\mn@eprint@arXiv#1{\href {http://arxiv.org/abs/#1} {{\tt arXiv:#1}}}
\def\mn@eprint@dblp#1{\href {http://dblp.uni-trier.de/rec/bibtex/#1.xml}
  {dblp:#1}}
\def\mn@eprint@#1:#2:#3:#4\@nil{\def\@tempa {#1}\def\@tempb {#2}\def\@tempc
  {#3}\ifx \@tempc \@empty \let \@tempc \@tempb \let \@tempb \@tempa \fi \ifx
  \@tempb \@empty \def\@tempb {arXiv}\fi \@ifundefined
  {mn@eprint@\@tempb}{\@tempb:\@tempc}{\expandafter \expandafter \csname
  mn@eprint@\@tempb\endcsname \expandafter{\@tempc}}}

\bibitem[\protect\citeauthoryear{Balbus}{Balbus}{1986}]{Balbus86}
Balbus S.~A.,  1986, \mn@doi [ApJL] {10.1086/184657}, 303, L79

\bibitem[\protect\citeauthoryear{{Blondin} \& {Koerwer}}{{Blondin} \&
  {Koerwer}}{1998}]{blondin98}
{Blondin} J.~M.,  {Koerwer} J.~F.,  1998, \mn@doi [\na]
  {10.1016/S1384-1076(98)00028-1}, \href
  {https://ui.adsabs.harvard.edu/abs/1998NewA....3..571B} {3, 571}

\bibitem[\protect\citeauthoryear{{Blondin} \& {Marks}}{{Blondin} \&
  {Marks}}{1996}]{BlondinMarks96}
{Blondin} J.~M.,  {Marks} B.~S.,  1996, \mn@doi [\na]
  {10.1016/S1384-1076(96)00019-X}, \href
  {https://ui.adsabs.harvard.edu/abs/1996NewA....1..235B} {1, 235}

\bibitem[\protect\citeauthoryear{{Blondin}, {Fryxell}  \& {Konigl}}{{Blondin}
  et~al.}{1990}]{blondin90}
{Blondin} J.,  {Fryxell} B.~A.,   {Konigl} A.,  1990, \apj, 360, 370

\bibitem[\protect\citeauthoryear{Carroll-Nellenback, Shroyer, Frank  \&
  Ding}{Carroll-Nellenback et~al.}{2013}]{carroll13}
Carroll-Nellenback J.~J.,  Shroyer B.,  Frank A.,   Ding C.,  2013, \mn@doi
  [Journal of Computational Physics]
  {http://dx.doi.org/10.1016/j.jcp.2012.10.004}, 236, 461

\bibitem[\protect\citeauthoryear{{Chevalier} \& {Imamura}}{{Chevalier} \&
  {Imamura}}{1982}]{CI}
{Chevalier} R.~A.,  {Imamura} J.~N.,  1982, \mn@doi [\apj] {10.1086/160364},
  \href {https://ui.adsabs.harvard.edu/abs/1982ApJ...261..543C} {261, 543}

\bibitem[\protect\citeauthoryear{{Ciardi} et~al.,}{{Ciardi}
  et~al.}{2009}]{Ciardi}
{Ciardi} A.,  et~al., 2009, \mn@doi [\apjl] {10.1088/0004-637X/691/2/L147},
  \href {https://ui.adsabs.harvard.edu/abs/2009ApJ...691L.147C} {691, L147}

\bibitem[\protect\citeauthoryear{{Cunningham}, {Frank}, {Varni{\`e}re},
  {Mitran}  \& {Jones}}{{Cunningham} et~al.}{2009}]{cunningham09}
{Cunningham} A.~J.,  {Frank} A.,  {Varni{\`e}re} P.,  {Mitran} S.,   {Jones}
  T.~W.,  2009, \mn@doi [\apjs] {10.1088/0067-0049/182/2/519}, \href
  {http://adsabs.harvard.edu/abs/2009ApJS..182..519C} {182, 519}

\bibitem[\protect\citeauthoryear{{Dgani}, {van Buren}  \&
  {Noriega-Crespo}}{{Dgani} et~al.}{1996}]{Dgani96}
{Dgani} R.,  {van Buren} D.,   {Noriega-Crespo} A.,  1996, \mn@doi [\apj]
  {10.1086/177114}, \href
  {https://ui.adsabs.harvard.edu/abs/1996ApJ...461..927D} {461, 927}

\bibitem[\protect\citeauthoryear{{Falle} \& {Raga}}{{Falle} \&
  {Raga}}{1993}]{Raga93}
{Falle} S.~A.~E.~G.,  {Raga} A.~C.,  1993, \mn@doi [\mnras]
  {10.1093/mnras/261.3.573}, \href
  {https://ui.adsabs.harvard.edu/abs/1993MNRAS.261..573F} {261, 573}

\bibitem[\protect\citeauthoryear{{Falle}, {Wareing}  \& {Pittard}}{{Falle}
  et~al.}{2020}]{Falle2020}
{Falle} S.~A.~E.~G.,  {Wareing} C.~J.,   {Pittard} J.~M.,  2020, \mn@doi
  [\mnras] {10.1093/mnras/staa131}, \href
  {https://ui.adsabs.harvard.edu/abs/2020MNRAS.492.4484F} {492, 4484}

\bibitem[\protect\citeauthoryear{Field}{Field}{1965}]{field65}
Field G.~B.,  1965, \mn@doi [\apj] {10.1086/148317}, 142, 531

\bibitem[\protect\citeauthoryear{{Folini} \& {Walder}}{{Folini} \&
  {Walder}}{2006}]{Folini06}
{Folini} D.,  {Walder} R.,  2006, \mn@doi [\aap] {10.1051/0004-6361:20053898},
  \href {https://ui.adsabs.harvard.edu/abs/2006A&A...459....1F} {459, 1}

\bibitem[\protect\citeauthoryear{{Frank} et~al.,}{{Frank}
  et~al.}{2014}]{Frankea2014}
{Frank} A.,  et~al., 2014, in {Beuther} H.,  {Klessen} R.~S.,  {Dullemond}
  C.~P.,   {Henning} T.,  eds, Protostars and Planets VI. p.~451 (\mn@eprint
  {arXiv} {1402.3553}), \mn@doi{10.2458/azu\_uapress\_9780816531240-ch020}

\bibitem[\protect\citeauthoryear{{Gardiner}, {Frank}, {Jones}  \&
  {Ryu}}{{Gardiner} et~al.}{2000}]{Gardiner00}
{Gardiner} T.~A.,  {Frank} A.,  {Jones} T.~W.,   {Ryu} D.,  2000, \mn@doi
  [\apj] {10.1086/308391}, \href
  {https://ui.adsabs.harvard.edu/abs/2000ApJ...530..834G} {530, 834}

\bibitem[\protect\citeauthoryear{{Gourdain}, Blesener, Greenly, Hammer, Knapp,
  Kusse  \& Schrafel}{{Gourdain} et~al.}{2010}]{gourdain10}
{Gourdain} P.~A.,  Blesener I.~C.,  Greenly J.~B.,  Hammer D.~A.,  Knapp P.~F.,
   Kusse B.~R.,   Schrafel P.~C.,  2010, \mn@doi [PhPl] {10.1063/1.3292653},
  17, 012706

\bibitem[\protect\citeauthoryear{{Hansen}, {Frank}, {Hartigan}  \&
  {Lebedev}}{{Hansen} et~al.}{2017}]{Hansen17}
{Hansen} E.~C.,  {Frank} A.,  {Hartigan} P.,   {Lebedev} S.~V.,  2017, \mn@doi
  [\apj] {10.3847/1538-4357/aa5ca8}, \href
  {https://ui.adsabs.harvard.edu/abs/2017ApJ...837..143H} {837, 143}

\bibitem[\protect\citeauthoryear{{Hartigan} et~al.,}{{Hartigan}
  et~al.}{2011}]{Hartigan11}
{Hartigan} P.,  et~al., 2011, \mn@doi [\apj] {10.1088/0004-637X/736/1/29},
  \href {https://ui.adsabs.harvard.edu/abs/2011ApJ...736...29H} {736, 29}

\bibitem[\protect\citeauthoryear{{Lamberts}, {Fromang}  \& {Dubus}}{{Lamberts}
  et~al.}{2011}]{Lamberts11}
{Lamberts} A.,  {Fromang} S.,   {Dubus} G.,  2011, \mn@doi [\mnras]
  {10.1111/j.1365-2966.2011.19653.x}, \href
  {https://ui.adsabs.harvard.edu/abs/2011MNRAS.418.2618L} {418, 2618}

\bibitem[\protect\citeauthoryear{{Langer}, {Chanmugam}  \& {Shaviv}}{{Langer}
  et~al.}{1981}]{Langer81}
{Langer} S.~H.,  {Chanmugam} G.,   {Shaviv} G.,  1981, \mn@doi [ApJL]
  {10.1086/183514}, 245, L23

\bibitem[\protect\citeauthoryear{{Lynden-Bell}}{{Lynden-Bell}}{1996}]{LyndenBell96}
{Lynden-Bell} D.,  1996, \mn@doi [\mnras] {10.1093/mnras/279.2.389}, 279,
  389–401

\bibitem[\protect\citeauthoryear{{McLeod} \& {Whitworth}}{{McLeod} \&
  {Whitworth}}{2013}]{McLeod13}
{McLeod} A.~D.,  {Whitworth} A.~P.,  2013, \mn@doi [\mnras]
  {10.1093/mnras/stt203}, 431, 710

\bibitem[\protect\citeauthoryear{{Mignone}}{{Mignone}}{2005}]{Mignone05}
{Mignone} A.,  2005, \mn@doi [\apj] {10.1086/429905}, \href
  {https://ui.adsabs.harvard.edu/abs/2005ApJ...626..373M} {626, 373}

\bibitem[\protect\citeauthoryear{{O'Sullivan} \& {Ray}}{{O'Sullivan} \&
  {Ray}}{2000}]{Osullivan00}
{O'Sullivan} S.,  {Ray} T.~P.,  2000, \aap, \href
  {https://ui.adsabs.harvard.edu/abs/2000A&A...363..355O} {363, 355}

\bibitem[\protect\citeauthoryear{{Pittard}, {Dobson}, {Durisen}, {Dyson},
  {Hartquist}  \& {O'Brien}}{{Pittard} et~al.}{2005}]{Pittard}
{Pittard} J.~M.,  {Dobson} M.~S.,  {Durisen} R.~H.,  {Dyson} J.~E.,
  {Hartquist} T.~W.,   {O'Brien} J.~T.,  2005, \mn@doi [\aap]
  {10.1051/0004-6361:20042260}, \href
  {https://ui.adsabs.harvard.edu/abs/2005A&A...438...11P} {438, 11}

\bibitem[\protect\citeauthoryear{{Ramachandran} \& {Smith}}{{Ramachandran} \&
  {Smith}}{2005}]{Ramachandran05}
{Ramachandran} B.,  {Smith} M.~D.,  2005, \mn@doi [\mnras]
  {10.1111/j.1365-2966.2005.08691.x}, \href
  {https://ui.adsabs.harvard.edu/abs/2005MNRAS.357..707R} {357, 707}

\bibitem[\protect\citeauthoryear{{Ramachandran} \& {Smith}}{{Ramachandran} \&
  {Smith}}{2006}]{Ramachandran06}
{Ramachandran} B.,  {Smith} M.~D.,  2006, \mn@doi [\mnras]
  {10.1111/j.1365-2966.2005.09890.x}, \href
  {https://ui.adsabs.harvard.edu/abs/2006MNRAS.366..586R} {366, 586}

\bibitem[\protect\citeauthoryear{{Ray}, {Dougados}, {Bacciotti},
  {Eisl{\"o}ffel}  \& {Chrysostomou}}{{Ray} et~al.}{2007}]{rayPPV2007}
{Ray} T.,  {Dougados} C.,  {Bacciotti} F.,  {Eisl{\"o}ffel} J.,
  {Chrysostomou} A.,  2007, in {Reipurth} B.,  {Jewitt} D.,   {Keil} K.,  eds,
  Protostars and Planets V. p.~231 (\mn@eprint {arXiv} {astro-ph/0605597})

\bibitem[\protect\citeauthoryear{{Ryutov}, {Drake}  \& {Remington}}{{Ryutov}
  et~al.}{2000}]{ryutov2000}
{Ryutov} D.~D.,  {Drake} R.~P.,   {Remington} B.~A.,  2000, \mn@doi [\apjs]
  {10.1086/313320}, 127, 465

\bibitem[\protect\citeauthoryear{{Steinberg} \& {Metzger}}{{Steinberg} \&
  {Metzger}}{2018}]{Steinberg18}
{Steinberg} E.,  {Metzger} B.~D.,  2018, \mn@doi [\mnras]
  {10.1093/mnras/sty1641}, \href
  {https://ui.adsabs.harvard.edu/abs/2018MNRAS.479..687S} {479, 687}

\bibitem[\protect\citeauthoryear{{Stevens}, {Blondin}  \& {Pollock}}{{Stevens}
  et~al.}{1992}]{Stevens92}
{Stevens} I.~R.,  {Blondin} J.~M.,   {Pollock} A.~M.~T.,  1992, \mn@doi [\apj]
  {10.1086/171013}, \href
  {https://ui.adsabs.harvard.edu/abs/1992ApJ...386..265S} {386, 265}

\bibitem[\protect\citeauthoryear{{Strickland} \& {Blondin}}{{Strickland} \&
  {Blondin}}{1995}]{stricklandBlondin95}
{Strickland} R.,  {Blondin} J.,  1995, \mn@doi [\apj] {10.1086/176093}, 449,
  727

\bibitem[\protect\citeauthoryear{{Sutherland}, {Bicknell}  \&
  {Dopita}}{{Sutherland} et~al.}{2003}]{sutherland03b}
{Sutherland} R.~S.,  {Bicknell} G.~V.,   {Dopita} M.~A.,  2003, \mn@doi [\apj]
  {10.1086/375294}, \href
  {https://ui.adsabs.harvard.edu/abs/2003ApJ...591..238S} {591, 238}

\bibitem[\protect\citeauthoryear{{Suzuki-Vidal} et~al.,}{{Suzuki-Vidal}
  et~al.}{2009}]{suzukiVidal09}
{Suzuki-Vidal} F.,  et~al., 2009, \mn@doi [\apss] {10.1007/s10509-009-9981-1},
  322, 19

\bibitem[\protect\citeauthoryear{{Suzuki-Vidal} et~al.,}{{Suzuki-Vidal}
  et~al.}{2012}]{suzukiVidal12}
{Suzuki-Vidal} F.,  et~al., 2012, \mn@doi [PhPl] {10.1063/1.3685607}, 19

\bibitem[\protect\citeauthoryear{{Suzuki-Vidal} et~al.,}{{Suzuki-Vidal}
  et~al.}{2015}]{suzukiVidal15}
{Suzuki-Vidal} F.,  et~al., 2015, \mn@doi [\apj] {10.1088/0004-637X/815/2/96},
  815:96

\bibitem[\protect\citeauthoryear{{Vishniac}}{{Vishniac}}{1994}]{Vishniac94}
{Vishniac} E.~T.,  1994, \mn@doi [ApJ] {10.1086/174231}, 528, 186

\bibitem[\protect\citeauthoryear{{Walder} \& {Folini}}{{Walder} \&
  {Folini}}{1998}]{Walder98}
{Walder} R.,  {Folini} D.,  1998, \aap, \href
  {https://ui.adsabs.harvard.edu/abs/1998A&A...330L..21W} {330, L21}

\bibitem[\protect\citeauthoryear{{de Gouveia dal Pino} \& {Benz}}{{de Gouveia
  dal Pino} \& {Benz}}{1993}]{DalPino93}
{de Gouveia dal Pino} E.~M.,  {Benz} W.,  1993, \mn@doi [\apj]
  {10.1086/172785}, 410, 686

\bibitem[\protect\citeauthoryear{{de Gouveia dal Pino} \& {Birkinshaw}}{{de
  Gouveia dal Pino} \& {Birkinshaw}}{1996}]{dalPino96}
{de Gouveia dal Pino} E.~M.,  {Birkinshaw} M.,  1996, \mn@doi [\apj]
  {10.1086/178011}, \href
  {https://ui.adsabs.harvard.edu/abs/1996ApJ...471..832D} {471, 832}

\makeatother
\end{thebibliography}

\bsp

\label{lastpage}

\end{document}